\documentclass[11pt]{article}

\usepackage{amsthm,amscd,amssymb}

\usepackage{amsmath}
\usepackage{amsfonts}
\usepackage{graphicx}
\usepackage{psfrag}

\newcommand{\bm}[1]{\mbox{\boldmath $#1$}}

\DeclareFontFamily{OT1}{rsfs}{}
\DeclareFontShape{OT1}{rsfs}{m}{n}{ <-7> rsfs5 <7-10> rsfs7 <10->
rsfs10}{} \DeclareMathAlphabet{\mycal}{OT1}{rsfs}{m}{n}

\def\defi{\stackrel{\mbox{\tiny \bf def}}{=}}

\def\scri{{\mycal I}}
\def\Id{\mbox{\bm{Id}}}

\newtheorem{Tma}{Theorem}

\newtheorem{Def}{Definition}
\newtheorem{Lem}{Lemma}
\newtheorem{Cor}{Corollary}

\theoremstyle{remark} 
\def\tr{\mbox{tr}}
\def\riem{\mbox{Riem}}
\def \Mcal {(\mathcal{M}^{1,n+1},\eta)}
\def \Mcalfour {(\mathcal{M}^{1,3},\eta)}
\def \Kco  {K^{\Omega}}
\def \Kcom  {\bm{K^{\Omega}}}
\def \gamco {\gamma^\Omega}
\def \metesf {\overline{\gamma}}
\def \conesf {\overline{\nabla}} 
\def \volunit {\bm{\eta_{\mathbb{S}^n}}}
\def \volunitdos{\bm{\eta_{\mathbb{S}^2}}}
\def \vollambda{\bm{\eta_{S_\lambda}}}

\def\metesf{\overline{\gamma}}

\def \esfunit {\mathbb{S}^{n}}

\def\defi{\stackrel{\mbox{\tiny \bf def}}{=}}

\def\book#1#2#3#4#5#6#7#8{#1, ``#2'' in {\it #3}, #5 #6 (#7), #4 (#8).}

\def\JournalPrep#1#2#3{#1, ``#2'', #3.}
\def\Journal#1#2#3#4#5#6{#1, ``#2'', {\em #3} {\bf #4}, #5 (#6).}

\def\JDG{J. Diff. Geom.}
\def\CQG{Class. Quantum Grav.}
\def\JPA{\em J. Phys. A: Math. Gen.}
\def\PRD{Phys. Rev. {\bf D}}

\def\ANYAS{\em Ann. N. Y. Acad. Sci.}

\def\NC{\em Nuovo Cimento}
\def\AM{\em Ann. Math.}
\def\AdM{\em Advan. Math.}

\def\ATMP{\em Adv. Theor. Math. Phys.}

\def\AMM{\em Amer. Math. Monthly}
\def\EM{\em Elem. Math.}
\def\JHEP{\em J. High Energy Phys.}

\begin{document}



\title{On the Penrose inequality for dust null shells in the Minkowski
spacetime of arbitrary dimension}

\author{Marc Mars$^1$ and Alberto Soria$^2$ \\
Facultad de Ciencias, Universidad de Salamanca,\\
Plaza de la Merced s/n, 37008 Salamanca, Spain \\
$^1$ \,marc@usal.es,  $^2$ \,asoriam@usal.es, }

\maketitle


\begin{abstract}
A particular, yet relevant, particular case of the Penrose inequality 
involves null shells propagating in the Minkowski spacetime. Despite
previous claims in the literature, the validity of
this inequality remains open.
In this paper we rewrite this inequality 
in terms of the geometry of the surface obtained by
 intersecting the past null cone of the original surface $S$  with a
 constant time hyperplane and the ``time height'' function
of $S$ over this hyperplane.  We also specialize to the case when $S$
 lies in the past null cone of a point  and show the validity of
 the corresponding inequality in any dimension (in four dimensions
 this inequality was proved by Tod \cite{Tod1985}). Exploiting properties 
of convex hypersurfaces in Euclidean space we write down the 
Penrose inequality in the Minkowski spacetime  of
 arbitrary dimension $n+2$ as an inequality for two smooth
functions on the sphere $\mathbb{S}^n$. We finally obtain a sufficient
 condition for the validity of the Penrose inequality in the four
dimensional Minkowski spacetime and show that 
this condition is satisfied by a large class of surfaces.
\end{abstract}

\section{Introduction}

The Penrose inequality is a conjecture on spacetimes
containing specific types of spacelike co\-dimen\-sion-two
surfaces which play the role of quasi-local replacements of black holes.
In addition, the spacetime is assumed to satisfy an energy condition
and have suitable asymptotic behaviour at infinity. The
inequality bounds from below the total mass of the spacetime 
in terms of the area of the quasi-local black holes (or suitable
surfaces defined in terms of them). The Penrose
inequality is important because it provides
a strengthening of the positive mass theorem and also, and perhaps more
importantly, because its validity would give rather strong 
indirect support for the validity
of the weak cosmic censorship conjecture \cite{Penrose1969}.
In fact, the Penrose
inequality was originally put forward by Penrose \cite{Penrose1973} as
a way of identifying gravitational configurations that could violate
the weak cosmic censorship hypothesis. Since then, and given the absence
of counterexamples, the emphasis
has turned into trying to prove the conjecture. 
So far, the inequality has been proved
in full generality in the case of asymptotically flat spacetimes 
satisfying the null convergence condition and containing
a time-symmetric asymptotically flat spacelike hypersurface with
an inner boundary composed of outermost closed minimal surfaces. The case
of spacetime dimension four and connected inner boundary was dealt
with by Huisken and Ilmanen \cite{HuiskenIlmanen2001}. The case
of spacetime dimension up to eight and no assumption on connectedness
of the inner boundary is due to Bray \cite{Bray2001}. The other 
general case where the inequality is known to hold is for
spherically symmetric spacetimes 
satisfying the dominant energy condition (in arbitrary 
spacetime dimension) \cite{MalecMurchadha1994,Hayward1996}.
Besides this, there are also many partial results
of interest (see \cite{Mars2009} for a relatively recent review
on this topic and  \cite{Recent} for some new developments).

One version of the Penrose inequality deals with asymptotically flat
spacetimes with a regular past null infinity $\scri^{-}$ and
satisfying the dominant energy condition. Consider a closed, orientable spacelike $S$ surface and recall
that $S$ admits two future directed
null normals $l$, $k$. If $S$ is such that the null expansion along $l$
vanishes (i.e. it is a marginally outer trapped surface) and, moreover,
the null hypersurface $\Omega$ defined by null geodesics starting on $S$
and tangent to $-k$ extends smoothly all the way to $\scri^{-}$, 
then the Penrose inequality conjectures that the Bondi mass 
evaluated at the cut between $\Omega$ and $\scri^{-}$ is bounded below
in terms of the area of $S$. In four dimensions, the inequality reads
\begin{eqnarray}
M_B \geq \sqrt{\frac{|S|}{16 \pi} }.
\label{PIBondi}
\end{eqnarray}
Ludvigsen and Vickers \cite{LudvigsenVickers1983} 
proposed an argument to prove this inequality
in the general case. However, this argument 
made use of an implicit assumption that does not hold in general
\cite{Bergqvist1997}. Moreover, it is not easy to write down conditions
directly on $S$ which ensures that this extra assumption holds true. The
Penrose inequality for the Bondi mass is therefore an open and interesting
problem.

A particularly simple case of this version of the inequality can be formulated
for spacetimes generated by shells of null dust propagating in the 
Minkowski spacetime. In fact this situation was the original set up where
the Penrose inequality was discussed in his seminal paper  \cite{Penrose1973}.
The  idea is the following: imagine that a infinitesimally thin shell
of matter is sent from past null infinity in the Minkowski spacetime. The matter
content of the shell is null dust, i.e. pressure-less matter propagating at the 
speed of light. Assume that the shape of the shell, as seen
by an inertial observer in the Minkowski spacetime, is a convex
surface sufficiently near $\scri^{-}$. This guarantees that the null hypersurface
defined by the motion of the null dust is regular in a neighbourhood of $\scri^{-}$.
Of course this null hypersurface will develop singularities in the future, where incoming
null geodesics meet conjugate points.  We denote by $\Omega$ the maximal extension of this null hypersurface
as a smooth submanifold in the Minkowski spacetime. 

The shell modifies the spacetime geometry
after it has passed and, since it is collapsing, it 
will typically  generate a trapped surface in its exterior. The spacetime outside the shell
is, in general, very complicated (in particular, because
the shell produces gravitational
waves), but the interior geometry remains unaffected before the shell goes through. Now, the spacetime 
geometry right after the shell passes can be determined from the interior geometry and the properties
of the shell itself by using the junction conditions between spacetimes, see e.g.
\cite{MarsSenovilla1993}. Moreover, the matter distribution of the shell 
can be prescribed freely (at one instant of time).  It turns out that given {\it any} closed (i.e. compact
and without boundary), spacelike surface embedded in $\Omega$, the energy distribution of the null shell can
be arranged so that $S$ is a marginally outer trapped surface with respect to the spacetime geometry generated
by the shell. The direction $l$ along which the null expansion vanishes is transverse to $\Omega$. Moreover, the 
energy density of the shell determines the Bondi mass of the newly generated spacetime at the cut defined
by the intersection of $\Omega$ and $\scri^{-}$. Now, the area of $S$ is the same when measured with 
the Minkowskian geometry and when observed from the outside spacetime geometry. Moreover, the jump of 
the null expansion $\theta_l$ across the shell can be computed in terms of the energy-density of the shell. Combining this facts,
it follows that the Penrose inequality (\ref{PIBondi}) becomes an inequality for (a suitable class of) closed spacelike
surfaces in the Minkowski spacetime. The resulting inequality is 
(see \cite{Penrose1973, Tod1985, Tod1998} in four spacetime dimensions  and \cite{Gibbons1997} for arbitrary dimension),
\begin{equation}
\label{B2}
\int_S\theta_l {\bm{\eta_S}} 
\geq n(\omega_{n})^{\frac{1}{n}}|S|^{\frac{n-1}{n}},
\end{equation}
where $n$ is the dimension of $S$ (i.e. two in spacetime dimension four)
and $\omega_n$ the area of the standard sphere $\mathbb{S}^n$.
In this expression $|S|$ is the area of the surface $S$ and 
$\theta_{l}$ is the null expansion of $S$
with respect to the future directed, outer (i.e. transverse to $\Omega$) null normal $l$ normalized by the condition 
$\langle l, k \rangle = -2$, where $k$ is the future directed null normal to $S$ which is tangent to $\Omega$
and which satisfies $\langle \partial_t ,k \rangle  = -1$. All these expressions refer to the geometry of the Minkowski
spacetime, in particular $\langle \cdot , \cdot \rangle$ denotes scalar product with the Minkowskian metric and
$\partial_t$ is a covariantly constant, unit, timelike vector field in the Minkowski spacetime.
The only restriction on the surfaces $S$ is that 
the null hypersurface obtained by sending light
orthogonally from them along $-k$ generates a hypersurface which is regular everywhere  and
extends all the way to infinity. Geometrically, it is clear
that this occurs if and only if the intersection
of $\Omega$ with the constant hyperplane $\{ t = t_0 \}$ (for $t_0$
sufficiently negative) is a convex
surface of the Euclidean space. We call these surfaces $S$ {\it spacetime convex} in this paper.

Despite the simplicity of the ambient geometry, proving this inequality is still remarkably difficult.
The first case that was solved involved surfaces $S$ that 
lie on a constant time hyperplane $\{ t = t_0 \}$. In this case, Gibbons proved \cite{Gibbons1973,Gibbons1997}
that the inequality reduces to the classic Minkowski inequality
relating the total mean curvature and the area of convex surfaces
in Euclidean space (see expression (4) in Sect. 56 of \cite{BonnesenFenchel}). 

The second case refers
to surfaces $S$ contained in the past null cone of a point and leads to 
a non-trivial inequality for functions on the sphere
\cite{Penrose1973,BarrabesIsrael1991}. In spacetime dimension
four, its validity was proved 
by Tod \cite{Tod1985} using 
the Sobolev inequality in 
$\mathbb{R}^4$ applied to functions with suitable angular dependence.
Regarding the general
case, Gibbons claimed \cite{Gibbons1997} to have a general proof. However,
the argument contains a serious gap and the problem remains open.
This gap was first mentioned in \cite{Mars2009} without going into the details. In Section \ref{GibbonsSect} we discuss in more detail the argument used by
Gibbons and show where it fails.

Our main objective  in this paper is to express the Penrose inequality
in the Minkowski spacetime of arbitrary dimension in terms of the 
geometry of the convex
euclidean surface obtained by intersecting $\Omega$ with
a constant time hyperplane $\{t=t_0\}$ together with the time
height function $\tau = t |_S - t_0$. This is the contents of 
 Theorem \ref{T0}.
By applying a powerful Sobolev type inequality on the sphere due to Beckner
\cite{Beckner1993} we prove the validity of this inequality in the case when
the surface $S$ lies in the past null cone of a point (Theorem 
\ref{Tspherical}). This generalizes
to arbitrary dimension 
the result by Tod \cite{Tod1985} in spacetime dimension four mentioned above
and shows that a conjecture put forward by this author
regarding the optimal form of the inequality is in fact true. 
The geometry of convex, compact 
hypersurfaces in Euclidean space can be fully described in terms
of a single function $h$ on the unit sphere. This function is called
the ``support function'' and plays an important role in this paper.
In spacetime dimension four, the support function was already used in
\cite{Tod1992} in a related but different context. One of our
main results is
Theorem \ref{T1} where we write down the Penrose inequality in Minkowski 
as an inequality involving two smooth functions on the $n$-dimensional
sphere. Inspired by the argument by
Ludvigsen and Vickers \cite{LudvigsenVickers1983} 
and simplified later by Bergqvist \cite{Bergqvist1997}, we are able
to prove (Theorem \ref{tsolution})
the validity of this inequality in four spacetime
dimensions for a large class of surfaces.
This class contains a non-empty open set of surfaces. However, when applied to
surfaces lying on the past null cone of a point, the only case covered
by this theorem is when $S$ is a round sphere. Thus, the cases covered
by Theorem \ref{Tspherical} and by Theorem \ref{tsolution}
are essentially complementary, which 
indicates that any attempt of proving the Penrose inequality
in Minkowski in the general case will probably require a combination of both 
methods.

The plan of the paper is as follows. In section 2 we discuss Gibbons'
argument \cite{Gibbons1997} and explain in detail where it fails.  
For the sake of clarity, we use in this section the same 
notation and conventions of \cite{Gibbons1997}. 
In Section $\ref{notation}$ we
introduce the notation and conventions we use in this paper. We also 
recall some well-known facts on the geometry of null hypersurfaces used later.
In Section $\ref{S1}$ we
relate the two null expansions of a spacelike surface embedded
in a strictly static spacetime. Although applied in this paper only
in the case of the Minkowski spacetime this result is interesting 
on its own and has potential application for the Penrose inequality 
for null dust shells propagating in background spacetimes
more general than Minkowski.
In Section $\ref{SEC0}$ we introduce the notions of 
{\it spacetime convex null hypersurface} and
{\it spacetime convex surface} which are useful for stating  and studying the
Penrose inequality in Minkowski and we rewrite this inequality 
in terms of the geometry of the projected surface obtained by
intersecting the outer directed past null cone $\Omega$ of $S$ with a constant
time hyperplane $\{ t = t_0\}$, and the so-called ``height function'' 
$\tau$ which identifies $S$ within $\Omega$. 
In Section $\ref{termsupport}$ we introduce the support function $h$
for convex hypersurfaces in Euclidean space and rewrite
the inequality in terms of the  functions $\{h,\tau\}$ (and its derivatives),
as functions of the unit sphere $\mathbb{S}^n$. We show  that, in the particular
case when $\Omega$ is the past null cone of a point, the Penrose inequality 
follows from Beckner's inequality \cite{Beckner1993}.
In Section $\ref{dimfour}$, we
restrict ourselves to the four dimensional case and, exploit properties
of two-dimensional endomorphisms in order to simplify the inequality in terms
of $\{h,\tau\}$. The result is stated in Theorem \ref{T2}.
Finally, in Section \ref{SEC1} we prove the validity of
the inequality for a large class of surfaces in the four-dimensional 
case. The method of proof is inspired in the flow of surfaces
put forward by  Ludvigsen and Vickers 
\cite{LudvigsenVickers1983} and simplified and clarified
later by Bergqvist \cite{Bergqvist1997}. The explicit form we have 
for the inequality allows us make the method work for a much
larger class of surfaces than those covered by the original argument.
In future work we intend
to study whether this extension can be pushed from the Minkowski spacetime
discussed here to more general spacetimes with a complete past null infinity.

\section{A critical revision of Gibbons' argument}
\label{GibbonsSect}

In this section we discuss the gap in Gibbons' attempt \cite{Gibbons1997} to prove the general inequality (\ref{B2}).
Following the notation in
\cite{Gibbons1997}, we will denote by $T$ the spacelike, spacetime convex surface involved in the inequality. The future directed
null normals are called $n^{\alpha}$ and $l^{\alpha}$ and are chosen so that $n^{\alpha}$
is inward (i.e. the geodesics tangent to $n^{\alpha}$
generate the null hypersurface
$\Omega$ extending to $\scri^{-}$) and satisfy $n^{\alpha} t_{\alpha} = -1$ and $l^{\alpha} n_{\alpha} = -1$, where $t^{\alpha}$ is a covariantly
constant, unit, timelike vector field in Minkowski and indices are raised and lowered with the Minkowski metric $\eta_{\alpha\beta}$. 
Let us denote by $\nabla_{\alpha}$ 
the covariant derivative in the Minkowski spacetime.

The 
strategy in \cite{Gibbons1997} was to project $T$ along $t^{\alpha}$  onto a constant time hyperplane orthogonal to $t^{\alpha}$. The projected
surface is denoted by $\hat{T}$. The main idea was to rewrite (\ref{B2}) in terms of the geometry of $\hat{T}$ as a hypersurface
in Euclidean space. Gibbons finds that, whenever $\theta_l > 0$, the projected surface $\hat{T}$
has non-negative mean curvature and
its mean curvature (with respect to the outer unit normal tangent to the constant time hyperplane) 
$\hat{J}$, reads  (see expression (5.11) in \cite{Gibbons1997})
\begin{eqnarray}
\hat{J} = \sqrt{\frac{2}{\gamma}} \, \rho + \sqrt{2 \gamma} \, \mu, \label{hatj}
\end{eqnarray}
where $\gamma \defi -t^{\alpha} l_{\alpha}$, $2 \rho \defi 
\nabla_{\alpha} l^{\alpha}$ 
is the null expansion of $l^{\alpha}$ (hence $\rho = \frac{1}{4} \theta_l$ when compared with the normalization we used 
in (\ref{B2})) and $2 \mu \defi - \nabla_{\alpha} n^{\alpha} $ is minus the null expansion
along $n^{\alpha}$. 
As a consequence of (\ref{hatj}) and properties of the Minkowski spacetime
it follows
\begin{eqnarray}
\int_T \rho dA = \frac{1}{4} \int_{\hat{T}} \hat{J} d \hat{A},  
\label{Gibbons}
\end{eqnarray}
where $dA$, $d\hat{A}$ are, respectively,
the area elements of $T$ and $\hat{T}$. The area of $\hat{T}$ is not smaller that the area of $T$ and hence 
inequality (\ref{B2}) would follow from (\ref{Gibbons}) and the Minkowski-type inequality
\begin{eqnarray}
\int_{\hat{T}} \hat{J} d \hat{A}  
\geq n(\omega_{n})^{\frac{1}{n}}|\hat{T}|^{\frac{n-1}{n}}.
\label{meanconvex}
\end{eqnarray}
In 1994, Trudinger \cite{Trudinger1994} considered this inequality for general mean convex surfaces in Euclidean
space (i.e. surfaces
with non-negative mean curvature) and gave an argument for its proof using an elliptic method. However, 
according to Guan and Li \cite{GuanLi}, this argument turns out to be incomplete and the inequality is still open
(in \cite{GuanLi} a parabolic argument is proposed which proves the inequality for mean
convex {\it starshaped domains} in Euclidean space). Nevertheless, the main problem with Gibbons' argument does not lie in the validity
of (\ref{meanconvex})
but on the orthogonal projection leading to (\ref{hatj}). The projection is performed as follows. First extend $n^{\alpha}$ to an ingoing
null hypersurface ${\cal N}$ by solving the affinely parametrized null geodesic $n^{\alpha} \nabla_{\alpha} n^{\alpha} =0$ with initial data $n^{\alpha}$ on $S$. Similarly,
$l^{\alpha}$ is extended to a null vector field on the outgoing null hypersurface ${\cal L}$ passing through $S$ and with tangent vector
$l^{\alpha}$. These vector fields are then extended to a spacetime neighbourhood of $S$ by parallel transport along $t^{\alpha}$. With this
extension, we have $n^{\alpha} t_{\alpha} = -1$ everywhere. Defining $\gamma$ 
on this neighbourhood by 
$\gamma \defi -t^\alpha l_\alpha$, the following vector field can be introduced:
\begin{equation}
\hat{\nu}^\alpha=\frac{1}{\sqrt{2\gamma}}(l^\alpha-\gamma n^\alpha).
\label{nu}
\end{equation}
It follows immediately that $\hat{\nu}^{\alpha}$ is everywhere normal to $t^{\alpha}$. Morever, this field
is orthogonal to $\hat{T}$ and unit on this projected surface.
Gibbons used in $\cite{Gibbons1997}$ that the mean curvature $\hat{J}$ of the projected surface $\hat{T}$ can be
expressed as $\hat{J}=\nabla_\alpha\hat{v}^\alpha |_{\hat{T}}$. However, the definition of mean curvature gives
$\hat{J} = \nabla_\alpha\hat{v}^\alpha |_{\hat{T}} - \frac{1}{2}\hat{v}^\alpha\nabla_\alpha\langle\hat{v},\hat{v}\rangle 
|_{\hat{T}}$. Thus, the
expression used by Gibbons is only correct provided $\hat{v}^\alpha\nabla_\alpha \langle \hat{v},\hat{v} \rangle |_{\hat{T}} =0$. The extension 
of $\hat{\nu}^{\alpha}$ is uniquely fixed by the definition (\ref{nu}) and a priori there is no reason why
this vector should remain unit in a neighbourhood of $\hat{T}$ (or, more
precisely, that the derivative of its norm
should vanish on $\hat{T}$). Moreover, substituting
(\ref{nu}) in the (correct) expression for $\hat{J}$  gives
\begin{eqnarray}
\label{meancurv}
\hat{J} = \left . \sqrt{\frac{2}{\gamma}} \, \rho + \sqrt{2 \gamma} \, \mu + l^{\alpha} \nabla_{\alpha} \left (
\frac{1}{\sqrt{2 \gamma}} \right ) - n^{\alpha} \nabla_{\alpha} \left ( \sqrt{\frac{\gamma}{2}} \right )-\frac{1}{2}
\hat{v}^\alpha \nabla_\alpha\langle \hat{v}, \hat{v} \rangle \right |_{\hat{T}},
\end{eqnarray}
which agrees with (\ref{hatj}) only if the last three terms cancel each other. The third term in the right-hand side
of (\ref{meancurv}) is always zero because
%
$l^{\alpha} \nabla_{\alpha} \gamma = 
l^{\alpha} \nabla_{\alpha} \left ( - t^{\beta} l_{\beta} \right ) = - t^{\beta} l^{\alpha} \nabla_{\alpha} l_{\beta} =0$,
which follows from the fact that $l^{\alpha}$ is geodesic and  $t^{\alpha}$ is covariantly constant. However, neither
$\hat{v}^\alpha\nabla_\alpha \langle \hat{v} , \hat{v} \rangle$ nor the derivative
of $\gamma$ along $n^{\alpha}$ need to vanish on $\hat{T}$. Even more, they need not, and in fact do not, cancel
out in general.  This fact invalidates
(\ref{hatj}) which in turn, spoils the relationship (\ref{Gibbons}) between the left-hand side of the
Penrose inequality (\ref{B2})  and the integral of the mean curvature $\hat{J}$ of the projected surface ${\hat T}$.
It is possible to derive general 
expressions both for $n^{\alpha} \nabla_{\alpha} \gamma$  and for 
$\hat{v}^\alpha \nabla_\alpha\langle \hat{v},\hat{v} \rangle$ on $\hat{T}$ (or $T$)
which show that such cancellations do not occur.
Instead of doing so, we find it more convenient to present an explicit example where 
the last two terms in (\ref{meancurv}) do not cancel each other. For completeness, we also evaluate 
$\hat{J}, \rho$ and $\mu$ explicitly on this example and show that (\ref{hatj}) is not valid.

For the example, we consider spherical coordinates $\{ t, r, \theta, \phi \}$ on Minkowski and consider the past null cone
of the origin $p$ defined by the coordinates $\{ t=0, r = 0 \}$. This past null cone $\Omega_p$ is defined by the 
equation $t + r =0$. We consider an axially symmetric (with respect to the Killing vector $\partial_{\phi}$) spacelike surface
$T$ embedded in $\Omega_p$. The embedding is then given by $\{ t = -R (\theta), r = R (\theta), \theta, \phi \}$, where $R$ is a smooth,
positive function (satisfying suitable regularity properties at the north and south poles, as usual). With the normalization
for $n^{\alpha}$, $l^{\alpha}$  above 
(and choosing $t^{\alpha} = (\partial_t)^{\alpha})$ a direct calculation gives
\begin{eqnarray}
\vec{n} \, |_{T} &=&\partial_t-\partial_r, \\
\vec{l} \, |_{T} &=& \left (\frac{R^2+(R')^2}{2R^2}\right ) \partial_t
+ \left (\frac{R^2-(R')^2}{2R^2}\right )\partial_r
-  \frac{R'}{R^2} 
\partial_\theta, \label{onT}
\end{eqnarray}
where prime denotes derivative with respect to $\theta$. We need to determine the vector field
$l^{\alpha} (y^{\beta})$ as a function of the spacetime coordinates $y^{\beta} = \{ t, r, \theta, \phi \}$. The condition that $l^{\alpha}$ is
paralelly propagated along $t^{\alpha}$ means that $l^{\beta}$ does not depend of $t$, i.e. $l^{\beta} (y^i)$,
with $y^i = \{ r, \theta, \phi \}$. The boundary conditions (\ref{onT})  on $T$ 
require
\begin{eqnarray}
& l^t (r, \theta,\phi )  \Big{|}_{r=R(\theta)} = \frac{R^2+(R')^2}{2R^2}, \hspace{2cm}  &  
l^r (r, \theta,\phi )  \Big{|}_{r=R(\theta)} = \frac{R^2-(R')^2}{2R^2}, \nonumber \\
&  l^{\theta} (r,\theta, \phi) \Big{|}_{r=R(\theta)} = -  \frac{R'}{R^2}, \hspace{2cm}  & 
 l^{\phi} (r,\theta, \phi) \Big{|}_{r=R(\theta)} = 0.  \label{expre}
\end{eqnarray}
Since $\gamma = - t^{\alpha} l_{\alpha} =
l^{t}$ it follows $n^{\alpha} \nabla_{\alpha} \gamma = - \partial_r l^t$. Thus, in spherical coordinates, 
$n^{\alpha} \nabla_{\alpha} \gamma |_{\hat{T}} = 
(\partial_r l^t) |_{r = R(\theta)}$. The component $\beta = 0$ of the geodesic equation $l^{\alpha} \nabla_{\alpha} l^{\beta} =0$
takes the explicit form 
\begin{eqnarray*}
l^{r}  \partial_r l^{t}  + l^\theta \partial_{\theta} l^t + l^{\phi} \partial_{\phi} l^t = 0. 
\end{eqnarray*}
Evaluating this expression on $r = R(\theta)$ and using (\ref{expre}) it is now straightforward to obtain
\begin{eqnarray}
n^{\alpha} \nabla_{\alpha} \gamma \, |_{\hat{T}} = 
- (\partial_r l^t) |_{r = R(\theta)} = 
\frac{- 2 (R')^2 \left ( R'' R - (R')^2 \right )}{R^3  \left ( R^2 + (R')^2 \right ) }. \label{third}
\end{eqnarray}
Applying a similar argument it follows that the last term of $\eqref{meancurv}$ takes the form
\begin{equation}
\left . \frac{1}{2}\hat{v}^\alpha\nabla_\alpha\langle\hat{v},\hat{v}\rangle \right |_{r=R(\theta)}=\frac{(R')^2}{R^2\sqrt{R^2+(R')^2}}.
\label{fourth}
\end{equation}
It is clear that (\ref{third}) and (\ref{fourth}) do not cancel each other in general. As mentioned above we complete
the argument by writing down the explicit expressions for $\gamma$, $\rho$, $\mu$ and $\hat{J}$.
It is a matter of simple calculation to obtain
\begin{eqnarray}
\gamma\, |_{\hat{T}} &=&-\langle
t,l\rangle|_{\hat{T}}=\frac{R^2+(R')^2}{2R^2}, \nonumber \\
\rho|_{\hat{T}}&=&
\frac{R^2+(R')^2-RR''}{2 R^3 }-\frac{R'}{2 R^2}\frac{\cos\theta}{\sin\theta}, \nonumber \\
\mu|_{\hat{T}}&=&\frac{1}{R}, \nonumber \\
\label{meanval}
\hat{J}|_{\hat{T}}&=&\frac{1}{\sqrt{R^2+(R')^2}}\left(
\frac{2 R^2+3(R')^2-RR''}{R^2+(R')^2}-\frac{R'}{R}\frac{\cos\theta}{\sin\theta}
\right).
\end{eqnarray}
Substituting these expressions in the right-hand side of (\ref{hatj}) gives
\begin{equation*}
\left . \sqrt{\frac{2}{\gamma}} \, \rho + \sqrt{2 \gamma} \,
\mu \frac{}{} \right |_{\hat{T}} =\frac{1}{\sqrt{R^2+(R')^2}}\left(
\frac{2R^2+2(R')^2-RR''}{R^2}-\frac{R'\cos\theta}{R\sin\theta}
\right),
\end{equation*}
which is clearly different to the expression for $\hat{J} |_{\hat{T}}$ in
(\ref{meanval}). This proves that (\ref{hatj}) cannot be correct. 
If we instead perform the analogous substitution in (\ref{meancurv})
we find a consistent expression.

\section{Notation and basic definitions}
\label{notation}
Throughout this paper $(M,g)$ denotes an  $(n+2)$-dimensional
spacetime, namely an $(n+2)$-dimensional oriented manifold endowed
with a metric $g$ of Lorentzian signature $+n$.  We always take $n
\geq 2$ and assume $(M,g)$ to be time oriented.  Tensors in $M$ carry
Greek indices and we denote by $\nabla$ the Levi-Civita covariant
derivative of $M$. On a manifold with metric $\gamma$, we denote by
$\langle \cdot, \cdot \rangle_{\gamma}$ the scalar product with this
metric. When the scalar product is with respect to the spacetime
metric $g$ we simply write $\langle \cdot, \cdot \rangle$. Our sign
convention for the Riemann tensor is $\riem(X,Y) Z \defi ( \nabla_X
\nabla_Y - \nabla_Y \nabla_X - \nabla_{[X,Y]} ) Z$.

The Penrose inequality in the Minkowski spacetime will involve
the geometry of null hypersurfaces, namely
codimension one, embedded submanifolds with degenerate first fundamental form.
Let $\Omega$ be a null hypersurface and $k$ a future directed
vector field tangent to $\Omega$ which is nowhere zero and null.
This vector field is defined up to multiplication with a positive
function $F : \Omega \rightarrow \mathbb{R}^+$. 
It is well-known
(see e.g. \cite{Galloway2000}) that given any point $p \in \Omega$, 
an equivalence relation can be defined on $T_p \Omega$ by means of
$X\sim Y$ iff $X-Y= c k$ with $c \in\mathbb{R}$.
The equivalence class of $X \in T_p \Omega$ is denoted by
$\bar{X}$ and the quotient space by $T_p \Omega / k$. 
The set  $T\Omega/k=\bigcup_{p\in\Omega}T_{p}\Omega/k$,
is endowed naturally with the structure of a vector bundle
over $\Omega$ (with fibers of dimension $n$) which is
called {\it quotient bundle}.

Given $\bar{X},\bar{Y} \in T_p \Omega/k  $, it follows that
$\gamco(\bar{X},\bar{Y}) \defi \langle X,Y\rangle$ is a positive
definite metric on this quotient space.
The tensor $\Kco(\bar{X},\bar{Y}) \defi \langle\nabla_{X} k ,Y\rangle$
is well-defined (i.e. independent of the representatives $X,Y \in T_p \Omega$
of $\bar{X}, \bar{Y}$ and of the extension of $Y$ to a neighbourhood of $p$).
This tensor is symmetric and
plays the role of a second fundamental form on $\Omega$.  The
{\it Weingarten map}, which we denote by $\Kcom$, is the endomorphism
obtained from $\Kco$ by raising one index with the inverse of $\gamco$.
Finally, the trace  of $\Kcom$ is the null expansion $\theta_k$ of 
$\Omega$. Under a rescaling $k \longrightarrow   F k$, these tensors
transform as $\Kco \longrightarrow  F \Kco, \Kcom \longrightarrow F \Kcom$ 
and $\theta_{k} \longrightarrow  F \theta_k$.

A derivative of $T \Omega/k$ can be defined via 
$(\overline{X})'\defi \overline{ \nabla_k X}$. Again this derivative is
well-defined (i.e. independent of the representative
chosen in the definition). Note, however, that it does
depend on the choice of $k$. As usual, this derivative
is extended to tensors in $T\Omega/k$ with the Leibniz rule.
An important property of null hypersurfaces is that
the quotient metric $\gamco$, the
quotient extrinsic curvature $\Kco$ and the ambient geometry $(M,g)$
are related by the following equations (see e.g. \cite{Galloway2000}),
which are the analog in the
null case to the standard Gauss-Codazzi equations for non-degenerate
submanifolds,
\begin{eqnarray*}
& & (\gamco)^{\prime} = 2 \Kco, \\
& & (\Kcom)'+\Kcom \circ \Kcom + \bm{R}  - Q \Kcom
=0, \quad \quad (\mbox{Ricatti equation} )
\end{eqnarray*}
where $\Kcom\circ \Kcom$ is the composition of endomorphisms,
$\bm{R}(\overline{X}) \defi\overline{\riem(X,k)k}$ and $Q$
is defined by $\nabla_k k  = Q k$ (the integrals curves
of $k$ are necessarily null geodesics but the parameter along them need
not be affine).

In order to transform this system of equations into a system of ODE for
tensor components, let us  choose $k$ to be affinely
parametrized, i.e. satisfying $\nabla_k k =0$. Let us 
also select $n$ vector fields $X_A$ ($A,B,C =1, \cdots, n$)
tangent to 
$\Omega$ satisfying the properties (i) $[k,X_A]=0$ and (ii)
$\{ k |_p , X_A |_p \}$ is a basis of $T_p \Omega$ at one 
point $p \in \Omega$. 
Denote by $\alpha_p (\sigma)$ an affinely
parametrized null geodesic containing $p$ and with tangent vector $k$ 
(for later convenience we do not fix yet the origin of the affine parameter
$\sigma$). Then $\{ \bar{X}_A |_{\alpha_p (\sigma)} \}$ is a basis of
$T_{\alpha_p (\sigma) } \Omega /k$ and the tensor coefficients
$\gamco_{AB} (\sigma) $, $\Kco_{AB} (\sigma)$ of $\gamco |_{\alpha_p(\sigma)}$
and $ \Kco |_{\alpha_p (\sigma)}$  in this basis satisfy the
ODE
\begin{eqnarray}
& &\frac{d(\Kco)^{A}_{\, \,B} }{d\sigma}=
-(\Kco)^{A}_{\,\,\,C}(\Kco)^{C}_{\,\,\, B}-R^{A}_{\,\,\, B}, \nonumber \\
& &\frac{d(\gamco)_{AB}}{d\sigma}=2(\Kco)_{AB}, \label{Ricatti}
\end{eqnarray}
where $R^A_{\,\,\,B} $ are defined by $\bm{R} (\bar{X}_B ) = R^{A}_{\,\,\,B} \bar{X}_A$
and indices are lowered and raised with the metric $(\gamco)_{AB}$
and its inverse $(\gamco)^{AB}$.

\section{Relationship between two null curvatures of a spacelike surface
in a strictly static spacetime}
\label{S1}

In this paper a {\it spacelike surface} $S$ is a connected, codimension-two, 
spacelike, oriented and closed (i.e. compact and without boundary)
smooth, embedded submanifold in a spacetime $(M,g)$. 
Tensors
in $S$ will carry Latin capital
indices and the induced metric and connection on $S$ are denoted 
respectively by $\gamma$ and $D$.
Our conventions for the second fundamental form  and
the mean curvature  are
$K(X,Y)\defi -(\nabla_{X}Y)^\bot$ and
$H \defi \mbox{Tr}K$. Here $X,Y$ are tangent vectors to $S$, and
$\bot$ denotes the normal component to $S$. 
If $\nu$ is a vector field
orthogonal to $S$, the second fundamental form
along $\nu$ is
$K^{\nu}(X,Y)\defi \langle
\nu,K(X,Y)\rangle=-\langle\nu,\nabla_{X}Y\rangle$,
with $X,Y\in T_{p}S $. 

The normal bundle of $S$ is a Lorentzian vector bundle which admits a
null basis $\{l,k\}$ that we always take smooth, future
directed and normalized so that 
$\langle k, l\rangle=-2$.
The second fundamental form $K^l$ along $l$ is also called
null extrinsic curvature and its trace is the null expansion along $l$,
denoted by $\theta_l$. The same applies to the null direction $k$.
The mean curvature decomposes in the null basis $\{l,k \}$
as $H =-\frac{1}{2}(\theta_k l +\theta_l k )$.          

Let us now assume that $M$ is strictly static, i.e.
that it admits a Killing vector field $\xi$ which is
everywhere timelike and hypersurface orthogonal. We define a positive
function $V$ by $\langle \xi, \xi \rangle=- V^2$. The integrability
of $\bm{\xi}$ implies, locally, the existence of a smooth function $t$
such that $\xi_{\alpha} = - V^2 \nabla_{\alpha} t$. The following Lemma
shows that the null expansions of any spacelike surface $S$ in a strictly
static spacetime are not independent to each other. In the case
of the Minkowski spacetime this result was proved  in \cite{Gibbons1997}.

\begin{Lem}[{\bf Relationship between null extrinsic curvatures}]
\label{L1}
Let $(M,g)$ be an $(n+2)$-dimensional strictly static spacetime with
static Killing vector $\xi$. Let $S$ be a spacelike surface in
$(M,g)$. With the notation above, we have
\begin{equation}
-\langle \xi,k\rangle  K^{l}_{AB}-\langle \xi,l\rangle  K^{k}_{AB}-D_{A}(\hat{V}^2
D_{B}\hat{t})-D_{B}(\hat{V}^2 D_{A}\hat{t})=0, \label{relationship}
\end{equation}
where $\hat{V}$ and $\hat{t}$ are respectively, the restriction of $V$ 
and $t$ on $S$.
\end{Lem}
\begin{proof}
Since the relationship is local, it suffices to work on a
suitably small neighbourhood $U_p$ of a point $p \in S$. We choose
$U_p$ small enough so that $\xi_{\alpha} = -V^2 \nabla_{\alpha} t$ on $U_p$
and work on $U_p$ from now on. Let
$\{ X_A \}$ be a basis of the tangent space to $S$ on $U_p$. Decomposing
$\xi $ in the basis $\{l,k,X_A\}$,  we find 
\begin{equation}
\label{A2}
\xi|_S = - \frac{1}{2} \langle \xi,k\rangle l  
- \frac{1}{2} \langle \xi,l\rangle k
- \hat{V}^2 X^C  X_C(\hat{t}), 
\end{equation}
where $X^C$ is the vector field $X^C \defi \gamma^{CD} X_D$.
The Killing equation $\nabla_{\alpha} \xi_{\beta} + \nabla_{\beta} \xi_{\alpha} =0$
 implies, on $U_p$,
\begin{equation}
\label{A3.1}
\langle  X_B, \nabla_{X_A} \xi \rangle + 
\langle  X_A, \nabla_{X_B} \xi \rangle   = 0.
\end{equation}
Let us work out the first term.
Inserting the decomposition (\ref{A2}) and using the definition of null
extrinsic curvature $K^{l}_{AB}= \langle X_B, \nabla_{X_A} l \rangle $
(and similarly for $k$) it follows
\begin{equation}
\label{A8}
\langle X_B, \nabla_{X_A} \xi \rangle  =
-\frac{1}{2}\langle \xi,k\rangle K^{l}_{AB} 
-\frac{1}{2}\langle \xi,l\rangle K^{k}_{AB}
- \left \langle X_B, \nabla_{X_A} ( \hat{V}^2 X^C X_C(\hat{t}) ) \right \rangle .
\end{equation}
Now, the tangential projection to $S$ of a spacetime covariant derivative
coincides with the intrinsic covariant derivative on $S$. More precisely,
for any vector fields $X,Y,Z$ tangent to $S$ we have
$\langle  X, \nabla_Y Z \rangle = \langle  X , D_Y Z \rangle_{\gamma}$.
Thus, the last term in (\ref{A8}) becomes
\begin{eqnarray*}
\left \langle X_B, \nabla_{X_A} ( \hat{V}^2 X^C X_C(\hat{t}) )
\right \rangle    &= & 
\left \langle X_B, D_{X_A} ( \hat{V}^2 X^C  X_C(\hat{t}) )
\right \rangle_{\gamma}  = 
\left \langle X_B, D_{X_A} X^C \right \rangle_{\gamma} 
\hat{V}^2 X_C ( \hat{t} )
 +  \\
 +X_A \left ( \hat{V}^2 X_B 
(\hat{t}) \right ) & = &  D_{A} ( \hat{V}^2 D_B  \hat{t}  ),
\end{eqnarray*}
where we used $\langle X_B, X^C \rangle_{\gamma} = \delta^C_{B}$
in the third equality and 
$\left \langle X_B, D_{X_A} X^C \right \rangle_{\gamma}  = - \Gamma^C_{BA}$,
where $\Gamma^C_{AB}$  are the connection coefficients of $D$ in the basis
$\{ X_A \}$. Inserting this expression into (\ref{A8}) we conclude
\begin{equation*}
\langle X_B, \nabla_{X_A} \xi \rangle  =
-\frac{1}{2}\langle \xi,k\rangle K^{l}_{AB} 
-\frac{1}{2}\langle \xi,l\rangle K^{k}_{AB}
- D_{A} ( \hat{V}^2 D_B  \hat{t}  )
\end{equation*}
which combined with (\ref{A3.1}) proves the Lemma.
\end{proof}

\begin{Cor}
\label{COR1}
Under the same assumptions as in the previous Lemma,
\begin{equation*}
\langle \xi,l\rangle \langle \xi,k\rangle =\hat{V}^2(1+\hat{V}^2|D\hat{t}|^2_\gamma).
\end{equation*}
\end{Cor}
\noindent (Here and in the following $|D f|^2_{\gamma} = \gamma^{AB} f_{,A} f_{,B}$  for any function $f: S \rightarrow \mathbb{R}$).
\begin{proof}
Squaring $\eqref{A2}$ it follows
\begin{eqnarray*}
- \hat{V}^2= \langle \xi, \xi \rangle = - \langle \xi,k\rangle \langle \xi,l\rangle + \hat{V}^4 \gamma^{CD}  X_C (\hat{t}) X_D (\hat{t}) 
=
- \langle \xi,k\rangle \langle \xi,l\rangle + \hat{V}^4  |D\hat{t}|^2_{\gamma}.
\end{eqnarray*}
\end{proof}

\section{Penrose inequality in the Minkowski spacetime in terms of the geometry of convex surfaces}
\label{SEC0}

We will restrict from now on to the $(n+2)$-dimensional Minkowski spacetime $\Mcal$ ($n \geq 2$). 
Choose a Minkowskian coordinate system $(t,x^\alpha)$  and define $\xi=\partial_t$. Since this Killing vector is unit, we have
$V =1$ in the notation of the previous section. The hyperplanes at constant $t = t_0$ will be denoted by
$\Sigma_{t_0}$.

 As already mentioned, the physical construction leading to
the Penrose inequality involves null hypersurfaces
which extend smoothly all the way to past null infinity. We introduce the following definition which
captures this notion conveniently (recall that a null hypersurface is {\it maximally extended} if it cannot be extended 
to a larger smooth null hypersurface).
\begin{Def}[{\bf Spacetime convex null hypersurface}]
Let $\Omega$ be a maximally extended null hypersurface in $\Mcal$. $\Omega$ is spacetime convex if there exists $t_0\in\mathbb{R}$
for which the surface $\widehat{S}_0=\Omega\cap \Sigma_{t_0}$ is closed (i.e. smooth, compact and without boundary),
connected and convex as a hypersurface of the euclidean geometry of $\Sigma_{t_0}$. $\Omega$ is called spacetime strictly convex if 
$\widehat{S}_0$ is strictly convex, namely with positive principal curvatures 
at every point.
\end{Def}
{\bf Remark.} The idea of the definition is, obviously, that if the shape of the null hypersurface at some instant of Minkowskian time is
convex, then the past directed outgoing null geodesics cannot develop caustics and hence the null hypersurface will extend smoothly to past null
infinity. It is also clear that if $\Omega\cap \Sigma_{t_0}$ is closed and convex for some $t_0$, the same occurs for all $t \leq t_0$.

\vspace{5mm}

Given a spacetime convex null hypersurface $\Omega$, we always normalize the tangent null vector $k$ uniquely by the condition
$\langle k, \xi \rangle = -1$. This vector field will also be normal to any spacelike surface embedded in $\Omega$. Since the Penrose
inequality involves precisely this type of surfaces the following definition is useful:

\begin{Def}[{\bf Spacetime convex surface}]
A spacelike surface $S$ embedded in $\Mcal$ is called 
spacetime (strictly) convex if it can be embedded in a spacetime (strictly)
convex null hypersurface $\Omega$ of 
$\Mcal$. 
\end{Def}
     
\begin{figure}[!htb]
\begin{center}
\psfrag{sig}{$\Sigma_{t_0}$}
\psfrag{s}{$S$}
\psfrag{t}{$\tau$}
\psfrag{k}{$k$}
\psfrag{l}{$l$}
\psfrag{so}{$\widehat{S}_0$}
\psfrag{kill}{$\xi$}
\psfrag{m}{$m$}
\psfrag{om}{$\Omega$}
\psfrag{esp}{$\Mcal$}
\includegraphics[width=10cm]{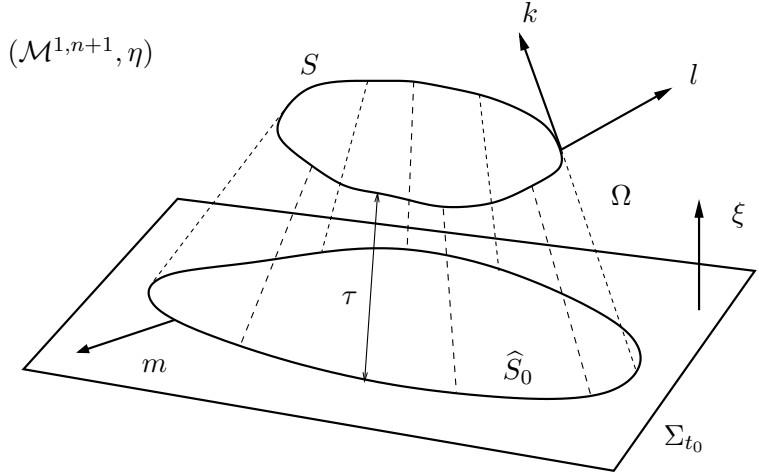}	
\end{center}
\caption{Schematic figure representing the 
construction above, where the spacetime
convex surface $S$ is projected along $\Omega$ onto the constant time hyperplane
$\Sigma_{t_0} = \{ t=t_0\}$.
The vectors on the normal bundle of $S$ are normalized
so that  $\langle k,\xi \rangle = -1$ and 
$\langle l, k \rangle = -2$. The vector field $m$ is unit, normal and 
pointing outside the surface $\widehat{S}_0$ within the hyperplane 
$\Sigma_{t_0}$. }
\label{fig1}
\end{figure}

It is intuitively obvious (and easy to prove)
 that a spacelike surface $S$ can be embedded at most in one spacetime convex null hypersurface $\Omega$.
Thus, for any such surface we can define unambiguously a null basis $\{l,k \}$ of its normal bundle by the conditions that $k$ is tangent to the
spacetime convex null hypersurface $\Omega$ containing $S$  and the normalization conditions $\langle k,\xi \rangle = -1$,
$\langle l, k \rangle = -2$.
We refer to $l$ as the outgoing null normal and to $k$ as ingoing null normal.
The Penrose inequality (\ref{B2}) involves the null expansion $\theta_l$ with respect to the outer null normal. The idea we want to explore
in this paper is how this inequality can be related to the geometry of a convex hypersurface of Euclidean space. The most natural convex surface arising in this setup is precisely the surface
 $\widehat{S}_0=\Omega\cap \Sigma_{t_0}$ (see Figure \ref{fig1}).
On the other hand, any convex surface 
$\widehat{S}_0 \hookrightarrow \Sigma_{t_0}$ defines uniquely a spacetime convex null hypersurface $\Omega$ and, then, any spacelike
surface embedded in $\Omega$ is defined uniquely by the ``time height'' function over $\Sigma_{t_0}$, namely the function $\tau \defi 
t |_S - t_0$. This function is defined on $S$. However, there is a canonical diffeomorphism $\phi : S \rightarrow \widehat{S}_{0}$ defined
by the condition that $\phi (p)$ lies on the maximally extended null geodesic $\alpha_p$ passing through $p$ and with tangent vector
$k |_p$. This diffeomorphism allows us to transfer geometric information from $S$ onto $\widehat{S}_0$ and viceversa. In particular, we can
define $(\phi^{-1} )^{\star} (\tau)$. Since no confusion will arise, we still denote this function by $\tau$. The precise meaning
will be clear from the context. 

\vspace{7mm}

\vspace{7mm}

The idea is thus to transform the Penrose inequality (\ref{B2}) into an inequality involving the geometry of $\widehat{S}_0$
as a hypersurface of Euclidean space $(\mathbb{R}^{n+1}, g_E)$ and the time height function $\tau$. The result is given in the
following theorem:

\begin{Tma}[{\bf Penrose inequality in Minkowski in terms of Euclidean geometry}]
\label{T0}
Let $\Mcal$ be the Minkowski spacetime with 
a selected Minkowskian coordinate system
$(t,x^{\alpha})$ $\xi = \partial_t$. Let $(S,\gamma)$ be a spacetime convex surface in $\Mcal$ and $\Omega$ the 
convex null hypersurface containing $S$. Consider a closed, convex surface $\widehat{S}_0 = \Omega \cap \Sigma_{t_0}$
as a hypersurface of Euclidean  space $(\mathbb{R}^{n+1}, g_E)$ and let $\gamma_0$ be its induced metric,
$\bm{\eta_{\widehat{S}_0}}$ its volume form, $K_0$ its second fundamental form with respect to the outer unit
normal and $\bm{K_0}$ the associated Weingarten map.
Then the Penrose inequality for  $S$ can be rewritten as
\begin{eqnarray}
& &
\int_{\widehat{S}_0} \left ( 1+
[(\Id-\tau \, \bm{K_0})^{-2}]^{A}_{\,\,\,C} (\gamma_{0}^{-1})^{CB}
\tau_{,A}\tau_{,B}\right ) \mbox{tr} 
\left [ \bm{K_0} \circ (\Id -\tau \, \bm{K_0})^{-1} \right ] \Delta[\tau]\bm{\eta_{\widehat{S}_0}}
\geqslant  
\nonumber \\
& &\geqslant 
n(\omega_n)^{\frac{1}{n}} \left ( \int_{\widehat{S}_0}\Delta[\tau]\bm{\eta_{\widehat{S}_0}} \right )^{\frac{n-1}{n}} 
\label{B1}
\end{eqnarray}
where $\Id$ is the identity endomorphism,
$\tau = t |_S - t_0$ and $\Delta[\tau]\defi \det(\Id -\tau \, \bm{K_0})$. 
\end{Tma}

\begin{proof}

Let us start by relating $\theta_l$ with $\theta_k$. Taking the trace of (\ref{relationship}) with respect to $\gamma$ (and using $\hat{V}=1$,
$\langle k,\xi\rangle = -1$):
\begin{eqnarray*}
 \theta_l-\langle \xi,l\rangle\theta_k-2\triangle_{\gamma}\tau = 0,
\end{eqnarray*}
where $\triangle_{\gamma} = D_A D^A$ is the Laplacian of $(S,\gamma)$.  Corollary 
$\eqref{COR1}$ gives $-\langle \xi,l\rangle=1+ |D\tau|_{\gamma}^2$  and the equation above becomes
\begin{equation*}
 \theta_l+(1+|D\tau|^2_{\gamma})\theta_k-2\triangle_\gamma\tau=0.
\end{equation*}
Integrating on $S$ it follows
\begin{equation}
\int_S\theta_l \bm{\eta_S}=-\int_S(1+|D\tau|^2_{\gamma})\theta_k\bm{\eta_S}, \label{thetalthetak}
\end{equation}
which gives the desired relationship.

The second step is to use the Ricatti equations on $\Omega$ in
order to relate $\theta_k$ on $S$ with the extrinsic geometry of $\widehat{S}_0$. To that aim, we first note that 
the vector field $k$ on $\Omega$ satisfies $\nabla_{k} k =0$ (this is an immediate consequence of the fact that 
$\xi$ is covariantly constant and $\langle \xi,k \rangle = -1$). Thus, the Ricatti equations on $\Omega$ take the form
(\ref{Ricatti}) provided we have selected 
$n$ vector fields $\{X_A \}$ tangent to $\Omega$ and satisfying the requirements that (i)  
$[k ,X_A] = 0 $ and (ii) $\{ k |_p , X_A |_p \}$ is a basis of $T_p \Omega, \forall p \in \Omega$
(more precisely $\{k, X_A \}$ is a basis of the tangent space of $\Omega$ on suitable
open subsets, however this abuse of notation is standard and poses no complications below).
Without loss
of generality we take $\{ X_A \}$ tangent to $S$. Equations (\ref{Ricatti}) still admit
the freedom of choosing the initial value of the affine
parameter $\sigma$ on each one of the null geodesics ruling $\Omega$. 
It turns out to be convenient to select
$\sigma$ so that $\sigma=0$ on $\widehat{S}_0$. This determines  $\sigma$ uniquely as a smooth function
$\sigma : \Omega \rightarrow \mathbb{R}$ which assigns to each point $p \in \Omega$,
the value of the affine parameter of the geodesic starting on $\widehat{S}_0$, with tangent vector $k$
and passing through $p$. Given that
\begin{eqnarray*}
k (t) = \bm{dt} \left ( k  \right ) = - \langle \xi, k \rangle = 1,
\end{eqnarray*}
and $t |_{\widehat{S}_0} = t_0$, it follows that $\sigma = t |_{\Omega} - t_0$. In 
particular $\sigma |_{S} = \tau$ (this is the main reason why this choice of the origin of the affine
parameter $\sigma$  is convenient). 

A crucial property of the geometry of a null hypersurface $\Omega$ is that, given 
any point $p \in \Omega$ and any embedded spacelike surface $S_p$ in $\Omega$ passing through $p$,
the induced metric $\gamma_{S_p}$ of $S_p$ and the second fundamental form $K^k_{S_p}$ of $S_p$ along the null
normal $k |_{p}$   satisfy $\gamma_{S_p} (X,Y) = \gamma^{\Omega} (\bar{X},\bar{Y} )$ and
$K^k_{S_p} (X,Y) = K^{\Omega} (\bar{X},\bar{Y} )$, where $X,Y \in T_p S_p$ (see e.g. \cite{Galloway2000}).
In other words, the induced metric and the extrinsic
geometry along $k$ of {\it any} embedded spacelike surface in $\Omega$ depends only on $p$ and not on the 
details of how $S_p$ is embedded in $\Omega$. Applying this result on $\widehat{S}_0$ we have, for any
point $\hat{p} \in \widehat{S}_0$,
\begin{eqnarray}
K^{\Omega} (\bar{X}_A,\bar{X}_{B} ) |_{\hat{p}} = K^{k}_{\widehat{S}_0} (\hat{X}_A , \hat{X}_B ) |_{\hat{p}} \label{first}
\end{eqnarray}
where $\hat{X}_A |_{\hat{p}}$ is defined by the properties (i) $\hat{X}_A |_{\hat{p}} \in \bar{X}_A |_{\hat{p}}$ and (ii)
$\hat{X}_A |_{\hat{p}}$ is tangent to $\widehat{S}_0$ at $\hat{p}$ (it is immediate that these two properties
define a unique $\hat{X}_A$). Now, the Jordan-Brouwer separation theorem (see e.g. \cite{Lima}) states that any 
connected, closed hypersurface
of Euclidean space separates $\mathbb{R}^n$ in two subsets, one with compact closure (called interior) 
and one with non-compact closure (called exterior).
Let $m$ be the unit normal of $\widehat{S}_0$ pointing towards the exterior, and denote by
$K_0$ the corresponding
second fundamental form and by $\bm{K_0}$ the associated Weingarten map.
Let $(K_0)_{AB}$ be the components of $K_0$ in the basis $\{ \hat{X}_A \}$.
Since $\Sigma_{t_0}$ is totally geodesic and $\langle k , m \rangle |_{\widehat{S}_0} = - 1$ (which follows from the fact
that $k$ is ingoing, future directed,  null  and satisfies $\langle k, \xi \rangle = -1$), we have
\begin{eqnarray}
K^{k}_{\widehat{S}_0} (\hat{X}_A , \hat{X}_B ) |_{\hat{p}}  = - (K_0)_{AB} |_{\hat{p}} \label{second}
\end{eqnarray}
Expressions (\ref{first}) and (\ref{second})  provides us with initial data 
$K^{\Omega}_{AB} |_{\sigma =0} = - (K_0)_{AB}$ 
for the Ricatti equation (\ref{Ricatti}), which in 
the Minkowski spacetime simplifies to 
\begin{eqnarray}
& &\frac{d(\Kco)^{A}_{\, \,B} }{d\sigma}=
-(\Kco)^{A}_{\,\,\,C}(\Kco)^{C}_{\,\,\, B},  \nonumber \\
& &\frac{d(\gamco)_{AB}}{d\sigma}=2(\Kco)_{AB}. \label{Ricatti2}
\end{eqnarray}
As it is well-known (and in any case easy to verify) the solution to these equations with initial data
$K^{\Omega}_{AB} |_{\sigma =0} = - (K_0)_{AB}$ is 
\begin{eqnarray}
\label{RicattiSol1}
& &(K^\Omega)^{A}_{\,\,\,B} \Big{|}_p = - \left . (K_0)^{A}_{\,\,\,C} \right |_{\pi(p)}  [(\Id - \sigma (p)  \bm{K_0} |_{\pi(p)})^{-1}]^{C}_{\,\,\,B}   \\
\label{RicattiSol2}
& &(\gamma^\Omega )_{AB} \Big{|}_p = \left . (\gamma_{0} )_{AC} \right |_{\pi(p)} [(\Id - \sigma(p) \bm{K_0} |_{\pi(p)})^2]^{C}_{\,\,\,B},
\end{eqnarray}
where  $\pi(p)$ is defined as the unique point on $\widehat{S}_0$ lying on the null geodesic $\alpha_p$. 
Now, the null expansion $\theta_k$  is related to $\Kco$ by
\begin{equation*}
\theta_k=\tr_{\gamma} K^{k}=\gamma^{AB}\langle\nabla_{X_A}k,X_B\rangle=\gamco(\overline{\nabla_{X_A}k},\overline{X_B})=(\gamco)^{AB}\Kco_{AB}=
(\Kco)^A_{\,\,A}.
\end{equation*}
Evaluating (\ref{RicattiSol1}) on $S$ (i.e. on $\sigma = \tau$) and taking the trace we find
$\theta_k |_p   =   -(K_0)^{A}_{\,\,\,C}
[(\Id-\tau \bm{K_0})^{-1}]^{C}_{\,\,\,A}   |_{\pi(p)}$, or equivalently
\begin{eqnarray*}
\theta_k \circ \phi^{-1} =  - \tr \left [ \bm{K}_0 \circ \left ( \Id - \tau \bm{K_0} \right )^{-1} \right ],  \label{thetak}
\end{eqnarray*}
where $\phi \defi\pi |_{S}$ is the diffeomorphism between $S$ and $\widehat{S}_0$ introduced above. 
In order to simplify the notation we will from now on suppress all references to $\phi$ when transferring information
from $S$ to $\widehat{S}_0$ via this diffeomorphism.

The remaining steps are to relate the volume forms of $S$ and  $\widehat{S}_0$ and to determine
$|D \tau|^2_{\gamma}$ (which appears in (\ref{thetalthetak})). Both involve the metric $\gamma$ on $S$.
Evaluating (\ref{RicattiSol2}) on $S$ and using $\gamma_{AB} = \gamma^{\Omega}_{AB}$ it follows
\begin{eqnarray}
\gamma_{AB}  = (\gamma_{0})_{AC}[(\Id - \tau \bm{K_0})^2]^{C}_{\,\,\,B}. \label{metricup0}
\end{eqnarray}
By construction, $\gamma_{AB}$ is positive definite, and hence invertible. Obviously this places
restrictions on the range of variation of $\tau$ (which clearly come from the fact that $\Omega$ cannot
be extended arbitrarily to the future as a smooth hypersurface). The precise range of variation of $\tau$ will be 
discussed below. Since $\gamma_0$ is positive definite, it follows that  $\Id - \tau \bm{K_0}$ is also
invertible and
\begin{equation}
(\gamma^{-1})^{AB}  =[(\Id-\tau \bm{K_0})^{-2}]^{A}_{\,\,\,C} (\gamma_0^{-1})^{CB}, \label{metricup}
\end{equation}
which implies, in particular,
\begin{equation}
|D\tau|^2_{\gamma}=[(\Id-\tau \bm{K_0})^{-2}]^{A}_{\,\,\,C} (\gamma_0^{-1})^{CB}\tau_{,A}\tau_{,B}.
\label{Dtau}
\end{equation}
Taking determinants in (\ref{metricup0}) it follows that the volume forms of $S$ and 
$\widehat{S}_0$ are related by
\begin{equation}
\bm{\eta_S}  =\Delta[\tau]\bm{\eta_{\widehat{S}_0}}   \label{etas}
\end{equation}
where $\Delta[\tau]\defi \det(\Id -\tau \, \bm{K_0})$.  Inserting (\ref{thetak}), (\ref{Dtau})
and (\ref{etas}) into (\ref{thetalthetak}) we find
\begin{eqnarray}
\int_S\theta_l \bm{\eta_S}=
\int_{\widehat{S}_0} \left ( 1+[(\Id-\tau \, \bm{K_0})^{-2}]^{A}_{\,\,\,C}
(\gamma_{0}^{-1})^{CB}
\tau_{,A}\tau_{,B}\right ) \tr  
\left [ \bm{K_0} \circ (\Id -\tau \, \bm{K_0})^{-1} \right ] \Delta[\tau]\bm{\eta_{\widehat{S}_0}},
\label{intthetal}
\end{eqnarray}
and the  Penrose inequality (\ref{B2}) becomes
(\ref{B1}), as claimed.
\end{proof}

A natural question for Theorem \ref{T0} is 
what is the class of functions $\tau : \widehat{S}_0 \rightarrow \mathbb{R}$ for which 
inequality (\ref{B1}) is conjectured. By construction, this amounts to
knowing which is the range of variation of $\sigma$ in $\Omega$. 
Let $\{\kappa_1 , \cdots, \kappa_n \}$ be the eigenvalues of $\bm{K_0}$, i.e. the principal curvatures
of $\widehat{S}_0$ as a hypersurface in Euclidean space. $\widehat{S}_0$ being convex, all these
curvatures are non-negative, and at least one of them is different from zero (because $\widehat{S_0}$ is closed).
The eigenvalues of the endomorphism $\Id - \tau \bm{K_0}$ are obviously 
$\{ 1 - \tau \kappa_1, \cdots , 1 - \tau \kappa_n \}$. Hence, this endomorphism is invertible as long as $\tau$ satisfies the bound
\begin{equation}
\tau  < \frac{1}{\underset{1 \leq A \leq n}{\max} \{\kappa_A \}}. 
\label{bound}
\end{equation}
Thus, the Penrose inequality, as written in inequality (\ref{B1}), is
conjectured to  hold for arbitrary smooth functions $\tau :
\widehat{S}_0 \rightarrow \mathbb{R}$ satisfying the pointwise bound
(\ref{bound}).  Incidentally, this statement also means that the range
of variation of $\sigma$ on the  null geodesic within $\Omega$ passing
through $\hat{p} \in \widehat{S}_0$ is $\sigma \in \left (-\infty,
\frac{1}{\max\{\kappa_A |_{\hat{p}} \} } \right )$.

\section{The Penrose inequality in terms of the support function}
\label{termsupport}
A remarkable property of convex hypersurfaces embedded in Euclidean space is
that a single function determines all of its geometric properties, both intrinsic
and extrinsic, in a very neat way. This function is called support function and is defined as follows:

\begin{Def}[{\bf Support function}]
Let $\widehat{S}_0$ be a closed, convex and connected hypersurface embedded in the Euclidean
space $(\mathbb{R}^{n+1},g_E)$.  Let $x (p)$ be the position vector of 
$p\in \widehat{S}_0$. The support function
$h:\widehat{S}_{0}\rightarrow\mathbb{R}$ is defined by $h(p)=\langle x(p),
m(p)\rangle_{g_E}$  where $m(p)$ is the unit normal at $p$ pointing towards 
the exterior of $\widehat{S}_0$.
\end{Def}

Closed, convex and connected hypersurfaces in $(\mathbb{R}^{n+1},g_E)$ are
always topologically $\mathbb{S}^n$. Moreover, if the surface
is strictly convex the Gauss map 
$m : \widehat{S}_0 \rightarrow \mathbb{S}^n$ is a diffeomorphism. We
will restrict ourselves to the strictly convex case from now on. 
This entails no loss of generality for the Penrose inequality because any convex
surface $\widehat{S_0}$ can be approximated
by strictly convex surfaces (e.g. by mean curvature flow \cite{Huisken1984}).
Let us denote by $\metesf$ the pull-back on $\widehat{S}_0$ of the standard
metric on the $n$-sphere and $\conesf$ the corresponding connection. Then, 
the induced metric $\gamma_0$ and second fundamental form $K_0$ 
of $\widehat{S}_0 \hookrightarrow
\mathbb{R}^{n+1}$ can be written in terms of the support function as follows
(see e.g. \cite{Smoczyk} p. 6):
\begin{eqnarray}
(K_0)_{AB}&=&\conesf_A\conesf_B h+\metesf_{AB}h \label{K0} \\
(\gamma_0)_{AB}&=&(\metesf^{-1})^{CD}(K_0)_{AC}(K_0)_{BD}.  \label{g0}
\end{eqnarray}  
Combining these formulas 
with Theorem \ref{T0} it becomes possible to rewrite 
the Penrose inequality for dust null shells in Minkowski
as an inequality on the sphere involving two smooth 
functions, namely $\tau$ and $h$. In this section we obtain the explicit
form of this inequality. To that aim, it is convenient to introduce
the endomorphism $\bm{B}$ obtained by raising one index to $K_0$
with the spherical metric $\metesf$, i.e. $B_{\,\,\,B}^{A} \defi  (\metesf^{-1})^{AC} (K_0)_{CB}$.
It is immediate from
(\ref{g0}) that $\bm{B}$ is the inverse endomorphism of the Weingarten
map ${\bm K_0}$. 
Since $\widehat{S}_0$ is diffeomorphic to $\mathbb{S}^n$ via the Gauss map we can identify
both manifolds and we can think of $\metesf$, $h$, ${\bm B}$ etc. as objects defined
on $\mathbb{S}^n$. This applies in particular to the function $\tau : \widehat{S}_0
\rightarrow \mathbb{R}$. With this notation, we can now state and prove the following
theorem, which gives the Penrose inequality in Minkowski in terms of the support function.

\begin{Tma}[{\bf Penrose inequality in Minkowski in terms of the support function}]
\label{T1}
Let $(S,\gamma)$ be a spacetime strictly convex
surface in $\Mcal$. With the same notation as in Theorem \ref{T0}, 
let $h$ be the support function of $\widehat{S}_0$. Then the Penrose inequality 
takes the form
\begin{eqnarray}
& &\int_{\mathbb{S}^n}\left (
1+ [( \bm{B}-\tau \Id )^{-2}]^{A}_{\,\,\,C} (\metesf^{-1})^{CB}  \tau_{,A}\tau_{,B}\right )
\mbox{tr}[ (\bm{B} - \tau \Id)^{-1} ] \det( \bm{B} -\tau \Id )\volunit
\geqslant
  \nonumber \\ & &\geqslant n(\omega_n)^{\frac{1}{n}}\left
  (\int_{\mathbb{S}^n}\det(\bm{B} -\tau \Id )\volunit\right
  )^{\frac{n-1}{n}}  \label{B3}
\end{eqnarray}
where $\metesf$, $\conesf$, $\volunit$ are the standard metric, connection
and volume form on $\mathbb{S}^n$, 
\begin{eqnarray}
B^A_{\,\,\,B} \defi (\metesf^{-1})^{AC}  \conesf_{C} \conesf_{B} h + \delta^{A}_{\,\,\,B} h, 
\label{secFund}
\end{eqnarray}
$h : \mathbb{S}^n \rightarrow \mathbb{R}$ is the support function of $\widehat{S}_0 
\hookrightarrow \mathbb{R}^{n+1}$
and $\tau : \mathbb{S}^n \rightarrow \mathbb{R}$ is the time height function of 
$S$.
\end{Tma}

\begin{proof} From (\ref{g0}) it follows that ${\bm B}$ determines the metric $\gamma_0$ via 
\begin{eqnarray}
(\gamma_0)_{AB}=B_{\,\,\,A}^{C}B_{\,\,\,B}^{D}\metesf_{CD} \label{g0bis}
\end{eqnarray}
which implies
\begin{equation}
\label{A10}
\bm{\eta_{\widehat{S}_0}}=\det( \bm{B})\volunit. 
\end{equation}
Since $\bm{B}$ is the inverse of $\bm{K_0}$, we have
\begin{eqnarray}
\Delta[\tau] \bm{\eta_{\widehat{S}_0}} = 
\det \left ( \Id - \tau \bm{K_0} \right ) \bm{\eta_{\widehat{S}_0}} = 
\det \left ( \Id - \tau \bm{K_0} \right ) \det (\bm{B}) \volunit = 
\det \left ( \bm{B} - \tau \Id \right ) \volunit.
\label{Deltaeta}
\end{eqnarray}
Similarly, 
\begin{eqnarray}
\tr  [ \bm{K_0} \circ \left (\Id - \tau \bm{K_0} \right )^{-1}  ]
= \tr  [ \bm{B^{-1}} \circ \left (\Id - \tau \bm{B^{-1}} \right )^{-1}  ] =
\tr  [ \left ( \bm{B} -  \tau \Id  \right )^{-1}  ].
\label{thetakbis}
\end{eqnarray}
It only remains to calculate 
$[(\Id-\tau \, \bm{K_0})^{-2}]^{A}_{\,\,\,C} (\gamma_{0}^{-1})^{CB}$.  From (\ref{g0bis})
and using again the fact that  $\bm{B}$ is the inverse of $\bm{K_0}$ we get
\begin{eqnarray}
& & [(\Id-\tau \, \bm{K_0})^{-2}]^{A}_{\,\,\,C} (\gamma_{0}^{-1})^{CB} =
[(\Id-\tau \, \bm{K_0})^{-2}]^{A}_{\,\,\,C} (K_0)^{C}_{\,\,\,D} (K_0)^{B}_{\,\,\,  F} 
(\metesf^{-1})^{DF} = \nonumber \\
& & = [(\Id-\tau \, \bm{K_0})^{-2}]^{A}_{\,\,\,C} (K_0)^{C}_{\,\,\,D} (K_0)^{D}_{\,\,\,  F} 
(\metesf^{-1})^{BF}.   \label{K1}
\end{eqnarray}
where in the last equality we made use of the property that $(K_0)^B_{\,\,\,F} (\metesf^{-1})^{DF}$
is symmetric (this follows from (\ref{g0}), which states in particular
that this tensor is the inverse of the symmetric two-covariant tensor
$(K_{0})_{BD}$). 
Since $(\Id - \tau \, \bm{K_0})^{-1} \circ {\bm K_0} = ( {\bm B} - \tau \, \Id )^{-1}$ it follows
\begin{eqnarray}
[(\Id-\tau \, \bm{K_0})^{-2}] \circ \bm{K_0} \circ \bm{K_0} =
[(\Id-\tau \, \bm{K_0})^{-1}] \circ ({\bm B} - \tau \, \Id)^{-1} \circ \bm{K_0} =
( {\bm B} - \tau \, \Id )^{-2} 
\label{KB}
\end{eqnarray}
where in the second equality we have used the fact that 
$(\Id-\tau {\bm K_0})$ and $({\bm B}-\tau \, \Id)$ commute. Using (\ref{KB}) in (\ref{K1}) yields
\begin{eqnarray}
[(\Id-\tau \, \bm{K_0})^{-2}]^{A}_{\,\,\,C} (\gamma_{0}^{-1})^{CB} =
[( {\bm B} - \tau \, \Id )^{-2} ]^A_{\,\,\,C} ( \metesf^{-1} )^{CB}. \label{new}
\end{eqnarray}
Substituting (\ref{Deltaeta}), (\ref{thetakbis}) and
(\ref{new}) into inequality (\ref{B1})  proves the theorem.
\end{proof}

In the following section we discuss the validity of the Penrose inequality for null 
dust shells in Minkowski when the incoming shell has spherical shape. Following 
\cite{Tod1992} we refer to this situation as the ``spherical case'' (note  however
that the incoming shell need not carry a spherically symmetric matter distribution). 
In other words, we consider the case when
the null hypersurface $\Omega$ is the past null cone of a point in Minkowski spacetime
and $S$ is any surface embedded in $\Omega$. The explicit form of this inequality in
spacetime dimension four appeared already in \cite{Penrose1973} and led to an inequality
for positive functions on the sphere. This inequality turned out to be highly non-trivial. 
Tod in \cite{Tod1985} was able to prove the inequality by using suitable functions on $\mathbb{R}^4$
and using the Sobolev inequality. In this paper we show that the Penrose inequality for spherical 
null dust shells in Minkowski holds in any spacetime dimension.

\subsection{Spherically symmetric case}

\begin{figure}[!htb]
\begin{center}
\psfrag{sig}{$\Sigma_{t_0}$}
\psfrag{s}{$S$}
\psfrag{t}{$\tau$}
\psfrag{k}{$k$}
\psfrag{l}{$l$}
\psfrag{so}{$\widehat{S}_0\simeq \mathbb{S}^n$}
\psfrag{kill}{$\xi$}
\psfrag{r}{$r$}
\psfrag{om}{$\Omega_p$}
\psfrag{esp}{$\Mcal$}
\psfrag{p}{$p$}
\includegraphics[width=10cm]{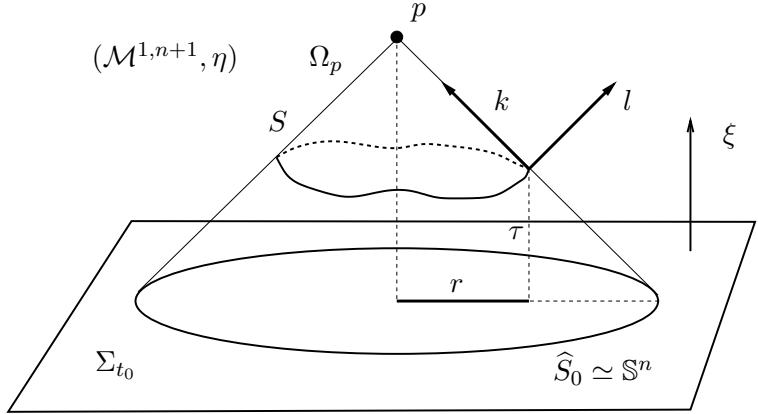}
\end{center}
\caption{When the spacetime convex surface $S$ lies in the past null cone of a point
$p\in\mathcal{M}^{1,n+1}$, its projection along $\Omega_p$ onto $\Sigma_{t_0}$ is always an n-sphere.
The Penrose inequality transforms in this case into an inequality for a single positive function $r$.}
\label{fig2}
\end{figure}

Let us restrict ourselves to the case when $\Omega$ is the past null
cone of a point (see Figure \ref{fig2}). As a consequence of Theorem \ref{T1}, 
the Penrose inequality transforms in
this case into an inequality for a single positive function on the sphere. Its
validity will follow as a simple consequence of the Beckner inequality
$\cite{Beckner1993}$ which bounds from above the $L^q$ norm of a
function on the sphere in terms of its $H^2$ norm. Specifically,

\begin{Tma}[{\bf Beckner, 1993}]
Let $F\in C^1(\esfunit)$ and denote as before 
the standard metric, volume form and connection of the $n$-dimensional unit sphere by
$\metesf$, $\volunit$, $\conesf$. Then
\begin{equation}
\frac{q-2}{n}\int_{\esfunit}|\conesf F|_{\metesf}^2 \volunit +\int_{\esfunit}|F|^2\volunit
\geq \left ( \omega_{n} \right )^{1- \frac{2}{q}} \left (\int_{\esfunit}|F|^q \volunit\right )^\frac{2}{q},
\label{beck} 
\end{equation}
where $2\leq q < \infty$ if $n=1$ or $n=2$ and
$2 \leq q \leq \frac{2n}{n-2}$ if $n\geq 3$. 
\end{Tma}

The following theorem settles the inequality when $\Omega$ is the past null cone of a point:
\begin{Tma}[{\bf Penrose inequality on a past null cone}]
\label{Tspherical}
Consider a point $p\in \mathcal{M}^{1,n+1}$ $(n\geqslant2)$
and $\Omega_p$ the past null cone of $p$. Let $S$ be a closed spacelike surface embedded
in $\Omega_p$. Then the Penrose inequality for $S$ reads
\begin{equation}
\int_{\esfunit}\left (r^{n-1}+r^{n-3}|\conesf r|^2_{\metesf}\right )\volunit\geq(\omega_{n})^{\frac{1}{n}}\left (\int_{\esfunit}r^{n}\volunit\right )^{\frac{n-1}{n}} 
\label{D3}
\end{equation}
where $r=t(p)-t|_S$. Moreover, this inequality holds true as a consequence of Beckner's theorem.
\end{Tma}

\begin{proof}
Select $t_0=t(p)-1$. Then, the function $\tau$ is written in terms of $r$
as $\tau=t|_S - t_0 = t|_S -t(p)+1=1-r$ and $\widehat{S}_0 = \Omega_p \cap \Sigma_{t_0}$ is
the $n$-dimensional unit sphere embedded in the Euclidean space. This surface
has support function $h = 1$, which implies
$(K_0)_{AB}= \conesf_A \conesf_B h + \metesf_{AB} h = \metesf_{AB}$ (this simply states
the well-known property that the unit sphere has
all principal curvatures equal to one). Then $B^A_{\,\,\,B}=(\metesf^{-1})^{AC}(K_0)_{CB}=\delta^A_{\,\,\,B}$
and  $(\bm{B} - \tau \Id ) =  ( 1 - \tau ) \Id = r \Id$, from which
\begin{eqnarray*}
1+[({\bf B}-\tau \Id)^{-2}]^A_{\,\,\,C}(\metesf^{-1})^{CB}\tau_{,A}\tau_{,B}&=&1+\frac{1}{r^2} |\conesf r|^2_{\metesf},
\\
\det \left ( {\bf B} - \tau \Id \right ) 
&=& r^n,  \\
\mbox{tr}[({\bf B}-\tau\Id)^{-1}] &=&  \frac{n}{r}.
\end{eqnarray*}
Substituting into (\ref{B3}) yields immediately (\ref{D3}). In order to show that this inequality is a particular
case of the Beckner inequality, we define $q=\frac{2n}{n-1}$ which clearly satisfies the bounds
$2 \leq q \leq \infty$ if $n=2$ and
$2 \leq q \leq \frac{2n}{n-2}$ if $n \geq 3$. Introducing the function
$F=r^{\frac{n-1}{2}}$, (\ref{D3}) becomes
\begin{equation}
\label{C2}
\left (\frac{2}{n-1}\right )^2\int_{\esfunit}|\conesf
F|^2_{\metesf} \volunit+\int_{\esfunit}F^2\volunit  
\geq
(\omega_n)^{1-\frac{2}{q}}\left (\int_{\esfunit}F^q\volunit\right
)^\frac{2}{q}.
\end{equation} 
Since $n\geq 2$, then
$\frac{q-2}{n}=\frac{2}{n(n-1)}\leq (\frac{2}{n-1})^2$ and inequality $\eqref{C2}$ is a particular case of $\eqref{beck}$.
\end{proof}

{\bf Remark.} As mentioned above, the case $n=2$ of this theorem was proved by Tod in \cite{Tod1985}
using the Sobolev inequality  in $\mathbb{R}^{4}$.  In a later paper, 
Tod proved \cite{Tod1986} that 
the factor $(\frac{2}{n-1})^2$ (i.e.
$4$ when $n=2$)  in front of the gradient in (\ref{C2}) could be improved to 
$8/3$ by using the Sobolev inequality of $\mathbb{R}^6$ applied to suitable functions. Tod
also conjectured that this factor could be improved to one. We note that 
Beckner's inequality implies in particular the validity of this conjecture by Tod.

\vspace{7mm}

\section{Penrose inequality in terms of the support function in spacetime dimension  four}
\label{dimfour}

The general expression for the Penrose inequality in terms of the support function as
written in Theorem \ref{T1} involves the inverse of the endomorphism ${\bf B} - \tau \Id$,
where  $B^{A}_{\,\,\,B} = \conesf^{A} \conesf_B h + \delta^{A}_{\,\,\,B}
h$ (for notational simplicity in this section we will lower and raise all indices with 
the spherical metric $\metesf$ and its inverse). Hence, the explicit form of the inequality
in terms of the support function is rather involved. In this section we restrict 
ourselves to spacetime dimension four, where the expressions simplify notably. The
reason is that, in this case, the endomorphism $\bm{B}$ acts on a two-dimensional vector space 
where inverses are much simpler to calculate. In fact, we will exploit the fact that
any endomorphism $\bm{A} : V_2 \rightarrow V_2$ acting on a two-dimensional vector
space  $V_2$ satisfies the identity 
\begin{eqnarray}
\bm{A}^2 = \tr(\bm{A}) \bm{A} - \det(\bm{A}) \Id. \label{Asquare}
\end{eqnarray}
This identity is a direct consequence of the expression of the minimal polynomial
in terms of the eigenvalues of $\bm{A}$ and the fact that these eigenvalues can be expressed
in terms of the trace and determinant of the endomorphism (alternatively, (\ref{Asquare})
can be proved by direct calculation in any basis). A simple consequence of (\ref{Asquare})
is that, whenever $\bm{A}$ is invertible
\begin{eqnarray}
\bm{A}^{-1}  = - \frac{1}{\det(\bm{A})} \bm{A}  + \frac{\tr(\bm{A})}{\det(\bm{A})} \Id. \label{inverse}
\end{eqnarray}
Taking traces in (\ref{Asquare}) and (\ref{inverse}) yields, respectively,
\begin{eqnarray}
\det (\bm{A}) & = &  \frac{1}{2} \left [ \tr (\bm{A})^2 - \tr(\bm{A}^2) \right ], \label{trace1} \\
\tr (\bm{A}^{-1} ) & = &  \frac{\tr (\bm{A})}{\det( \bm{A}) }. \label{trace2}
\end{eqnarray}
Squaring (\ref{inverse}) and using (\ref{Asquare}) and (\ref{trace1})
 we get an expression for $\bm{A}^{-2}$ which reads:
\begin{eqnarray}
\bm{A}^{-2} = - \frac{\tr (\bm{A})}{[\det(\bm{A})]^2}
\bm{A} + \frac{\left [\tr(\bm{A})^2 + \tr (\bm{A}^2 )  \right ]}{2 [\det(\bm{A})]^2} \Id.
\label{Aminus2}
\end{eqnarray}
Of particular interest below is the case when $A$ is of the form $\bm{A} = \bm{A_0} + f \Id$ for some scalar $f$.
Inserting this respectively into (\ref{trace1}) and (\ref{Aminus2}) gives, after a straightforward calculation,
\begin{eqnarray}
\det ( \bm{A_0} + f \Id ) &=&  \frac{1}{2} \left [ \tr (\bm{A_0})^2 - \tr(\bm{A_0^2}) \right ] + f \tr ( \bm{A_0}  ) + f^2 ,
\label{detdet} \\
\left ( \bm{A_0} + f \Id \right )^{-2} & =& 
- \frac{ \tr (\bm{A_0}) + 2 f }{[\det ( \bm{A_0} + f \Id )]^2} \bm{A_0} 
+ \frac{ \frac{1}{2} \left [  \tr (\bm{A_0})^2 + \tr ( \bm{A_0^2} ) \right ] + 2 f \tr (\bm{A_0}) + f^2
}{[\det ( \bm{A_0} + f \Id )]^2}  \Id. \label{Aminus2bis}
\end{eqnarray}
Having noticed these algebraic identities, we can now write down the specific form of the
Penrose inequality in terms of the support function in the case of four spacetime dimensions.
\begin{Tma}
\label{T2}
Let $(S,\gamma)$ be a spacetime strictly convex
surface in the Minkowski spacetime
$(\mathcal{M}^{1,3}, \eta)$. With the same notation as in 
Theorem $\ref{T1}$, the Penrose inequality can be written in the form
%
\begin{eqnarray}
& &
\int_{\mathbb{S}^2}\left (1+ W_1 | \conesf \tau|^2_{\metesf} - W_2 
(\conesf^A \conesf^B h) \conesf_A \tau \conesf_B \tau  \right )
\left ( \frac{}{}  \triangle_{\metesf}
  h+2(h-\tau) \right ) \volunitdos \geqslant \nonumber \\ 
& &\geqslant
\sqrt{16 \pi \int_{\mathbb{S}^2}\left (
(h-\tau)^2+(\triangle_{\metesf} h)(h-\tau)-\frac{1}{2}(h\triangle_{\metesf} h)
\right  )\volunitdos}
\label{B4}
\end{eqnarray}
where $\triangle_{\metesf}$ is the Laplacian of the unit  2-sphere and
\begin{eqnarray}
W_1 & \defi & \frac{ (h - \tau)^2 + 2( h - \tau) \triangle_{\metesf} h + \frac{1}{2} \left [ 
\left ( \triangle_{\metesf} h  \right )^2 + (\conesf_C \conesf_D h)(\conesf^C \conesf^D h) \right ]}{
\left [ (h - \tau)^2+ (h-\tau) \triangle_{\metesf} h+\frac{1}{2}[(\triangle_{\metesf}
  h)^2-(\conesf_C\conesf_D h)(\conesf^C\conesf^D h)] \right ]^{2}}, \label{W1}\\
W_2 & \defi & 
\frac{\triangle_{\metesf} h + 2 ( h- \tau) }{
\left [ (h - \tau)^2+ (h-\tau) \triangle_{\metesf} h+\frac{1}{2}[(\triangle_{\metesf}
  h)^2-(\conesf_C\conesf_D h)(\conesf^C\conesf^D h)] \right ]^2}. \label{W2}
\end{eqnarray}
\end{Tma}

\begin{proof}
Define the endomorphism $A_0{}^{A}_{\,\,\,B} = \conesf^A \conesf_B h$,  so that
$\tr (\bm{A_0}) = \triangle_{\metesf} h$ and $\bm{B} - \tau \Id = \bm{A_0} + ( h- \tau ) \Id$.
Applying identity (\ref{detdet}) with $f = h - \tau$ gives
\begin{eqnarray}
\det ( \bm{B} - \tau \Id ) = (h-\tau)^2+\triangle_{\metesf} h(h-\tau)+\frac{1}{2} \left [(\triangle_{\metesf}
  h)^2-(\conesf_C\conesf_D h)(\conesf^C\conesf^D h) \right ]. \label{first_a}
\end{eqnarray}
 Using (\ref{trace2}) we have
\begin{eqnarray}
\tr [( \bm{B} - \tau \Id)^{-1} ] \det ( \bm{B} - \tau \Id  ) = 
\tr ( \bm{B} - \tau \Id ) = \triangle_{\metesf} h + 2 (h- \tau).  \label{second_a}
\end{eqnarray}
We still need to evaluate $(\bm{B} - \tau \Id)^{-2}$ from (\ref{Aminus2bis}).  Using the definitions of 
$W_1$ and $W_2$ it is immediate to check that  
\begin{eqnarray}
(\bm{B} - \tau \Id)^{-2} = W_1 \bm{\Id} - W_2 \bm{A_0}.  \label{third_a}
\end{eqnarray}
Substituting (\ref{first_a}), (\ref{second_a}) and (\ref{third_a}) into the left-hand side
of inequality (\ref{B3}) gives the left-hand side of (\ref{B4}). In particular,
we have obtained an explicit formula for the integral of $\theta_l$ on $S$, namely
\begin{eqnarray}
\int_{S} \theta_l {\bm{\eta_S}} = 
\int_{\mathbb{S}^2}\left (1+ W_1 | \conesf \tau|^2_{\metesf} - W_2 
(\conesf^A \conesf^B h) \conesf_A \tau \conesf_B \tau  \right )
\left ( \frac{}{}  \triangle_{\metesf}
  h+2(h-\tau) \right ) \volunitdos. \label{Totalthetal}
\end{eqnarray}
For the right-hand side of (\ref{B4})
we need to calculate $|S|=\int_{\mathbb{S}^2}\det({\bf B}-\tau \Id)\volunitdos$. 
In particular, we need to integrate $( \triangle_{\metesf} h )^2 -
(\conesf_{C} \conesf_D h ) (\conesf^{C} \conesf^D h)$
on the sphere. We note the following identity
\begin{eqnarray}
 \conesf_C[(\conesf^C h)(\conesf_D\conesf^D h)]-\conesf_C[(\conesf^D
  h)(\conesf^C\conesf_D h)] & = &  \nonumber \\
=(\triangle_{\metesf} h)^2-(\conesf_C\conesf_D h)(\conesf^C\conesf^D h) 
+ \left ( \conesf^C h \right ) 
\left [ 
\conesf_C \conesf_D \conesf^D h - \conesf_D \conesf^D \conesf_C h \right ] & = & \nonumber \\
=(\triangle_{\metesf} h)^2-(\conesf_C\conesf_D h)(\conesf^C\conesf^D h) 
+ \left ( \conesf^C h \right ) 
\left [ 
\conesf_C \conesf_D \conesf^D h - \conesf_D \conesf_C \conesf^D h \right ] & = & \nonumber \\
 =(\triangle_{\metesf} h)^2-(\conesf_C\conesf_D h)(\conesf^C\conesf^D h) - 
| \conesf h |^2_{\metesf}  & & \label{Hess2}
\end{eqnarray}
where in the last equality we have used the definition of the Riemann tensor and the fact that
the sphere has constant curvature equal to one. Integrating (\ref{Hess2}) and using the fact that the left-hand side
of this expression is a divergence, it follows
\begin{equation}
\label{E3}
\int_{\mathbb{S}^2}\left ((\triangle_{\metesf} h)^2-(\conesf_C\conesf_D
h)(\conesf^C\conesf^D h) \right )\volunitdos=\int_{\mathbb{S}^2}
| \conesf h|^2_{\metesf} \volunitdos=\int_{\mathbb{S}^2}- (h\triangle_{\metesf}
h)\volunitdos
\end{equation}
where in the last step we have integrated by parts. Summing up,
\begin{equation}
\label{A11}
|S|=\int_{\mathbb{S}^2}\left ((h-\tau)^2+(\triangle_{\metesf}
h)(h-\tau)-\frac{1}{2}h\triangle_{\metesf} h\right )\volunitdos
\end{equation}
which inserted into the right-hand side of (\ref{B3}) gives the right-hand side of (\ref{B4}) (recall
that $\omega_2 = 4 \pi $).

\end{proof}

As already mentioned above, it is well-known that
when the surface $S$ lies in 
a hyperplane of the Minkowski spacetime, the Penrose inequality (\ref{B2})
becomes the classic Minkowski inequality for the total mean
curvature $\hat{J}$ of a surface in Euclidean space. In the case of $3+1$
dimensions, the Minkowski inequality reads
\begin{eqnarray}
\int_{\widehat{S}_0} \hat{J}  \bm{\eta_{\widehat{S}_0}} \geq \sqrt{16 \pi |\widehat{S}_0 |}.
\label{MinkIn}
\end{eqnarray}
Using the theorem above we can obtain the explicit form of the Minkowski inequality
in terms of the support function. This  result is obviously not new, but stated here for later
reference.

\begin{Cor}[{\bf Minkowski inequality in $(\bm{\mathbb{R}^3}, \bm{g_E})$
in terms of the support function}]
Let $\widehat{S}_0$ be a spacetime striclty convex surface embedded in a constant
time hyperplane of the Minkowski spacetime $\Mcalfour$. Then, the Minkowski
inequality (\ref{MinkIn}) in terms 
of the support function $h$ of $\widehat{S}_0$ takes the form
\begin{equation}
\label{minkineq}
\left
  (\int_{\mathbb{S}^2}h\volunitdos\right )\geqslant
  \sqrt{4\pi\int_{\mathbb{S}^2}\left ( h^2+\frac{1}{2}h\triangle_{\metesf}
  h \right ) \volunitdos }.
\end{equation} 
\end{Cor}

\begin{proof}
 Without loss of generality choose $t_0$ as the value of $t$ on the hyperplane where $\widehat{S}_0$ lies.
This choice implies $\tau =0$ and that $h$ is the support function of $\widehat{S}_0$. Since 
$\conesf_A \tau =0$,  inequality (\ref{B4}) reduces to (\ref{minkineq}).
\end{proof}

Inequality (\ref{B4}) in terms of the support function is still formidable.
However, it is completely explicit in terms of two functions on the sphere.
In the next section we prove its validity for a subset of admissible
functions $\{h,\tau\}$. This subset has not-empty interior (in any reasonable topology)
so the class  of surfaces where the inequality is proved is rather large.
The proof is inspired in the flow of surfaces
put forward by Ludvigsen and Vickers \cite{LudvigsenVickers1983} 
in their attempt to prove the general Penrose inequality in terms
of the Bondi mass. As mentioned in the Introduction, Bergqvist 
\cite{Bergqvist1997}
found a gap
in the argument and showed that the method provides a proof only 
under additional circumstances which are, in principle, not straightforward
to control directly in terms of the initial surface. In our situation
we have very explicit control of the whole flow of surfaces. This allows us, 
on the one hand, to find sufficient conditions for the validity of the Penrose inequality 
directly in terms of the geometry of the initial surface and, on the other, to prove the inequality 
for a much larger class than the one covered by Bergqvist's argument. In a future work we intend
to study in detail the relationship between the argument here and the
proof in \cite{Bergqvist1997} in order to see if the argument here admits
a generalization to general spacetimes with complete past null infinity.

\section{Dragging the surface along its past null cone}
\label{SEC1}

The flow put forward by Ludvigsen and Vickers 
\cite{LudvigsenVickers1983} 
and analyzed further by Bergqvist \cite{Bergqvist1997}
consists in dragging the initial surface $S$ along its outer directed
past null cone along affinely parametrized null geodesics. The key
property that makes this flow useful is the existence
of a monotonic quantity, often called Bergqvist mass. 
We start by introducing the flow
and defining the Bergqvist mass in our context.

We put ourselves in the setting where $S$ is a spacetime strictly convex
surface in the four-dimensional Minkowski spacetime $\Mcalfour$,
$\Omega$ is the spacetime convex null hypersurface where it sits
and $\widehat{S}_0 = \Omega \cap \Sigma_{t_0}$ is closed. We have introduced
in Section \ref{SEC0} a smooth function $\sigma : \Omega \rightarrow
\mathbb{R}$ which assigns to every point $p \in \Omega$,
the affine parameter at $p$ of the null geodesic tangent to the null vector
$k$ starting on $S$. By construction, $\sigma$ vanishes on $\widehat{S}_0$
and takes the values $\sigma |_S = \tau$. Let us extend $\tau$ to a function
$\tau : \Omega \rightarrow \mathbb{R}$ by imposing 
$k (\tau) =0$
and introduce a new smooth function $\tilde{\lambda} : \Omega \rightarrow
\mathbb{R}$ by $\tilde{\lambda} = \tau - \sigma$. Geometrically,
$\tilde{\lambda}$ is just a reparametrization of the null geodesics
ruling $\Omega$ (with this parameter the tangent vector is 
$-k$  and the geodesics start on $S$). It is immediate to see
that the level sets $S_{\lambda} = \{ \tilde{\lambda}^{-1} (\lambda), \lambda \geq 0 \}$
of this function define spacetime convex surfaces embedded
in $\Omega$. The collection of $\{S_{\lambda}  \}$, $\lambda \in [ 0, \infty)$
defines a flow starting at $S = S_0$. Let us denote by
$\gamma_{\lambda}$ and $\vollambda$ the induced metric and volume
form of $S_{\lambda}$ and by $\theta_l(\lambda)$ the outer null expansion 
of $S_{\lambda}$ (with the normalization $\langle l ,k \rangle = -2$,
as before). Then, the Bergqvist mass is defined by
\begin{eqnarray}
M_b (\lambda ) \defi \left ( \int_{S_\lambda}\theta_l(\lambda)\vollambda\right
) - 8 \pi \lambda. \label{Bergqvist}
\end{eqnarray}
In \cite{Bergqvist1997} the derivative of $M_b$ with respect to
$\lambda$ is calculated  using the spin formalism. For the sake
of completeness, let us rederive this derivative
using a purely tensorial formalism.
\begin{Lem}[{\bf Bergqvist \cite{Bergqvist1997}}]
\label{BergLem}
With the definitions above we have
\begin{eqnarray*}
\label{A14}
\frac{d M_b (\lambda)}{d \lambda} = 
-\int_{S_\lambda}2\langle s_{\lambda},s_{\lambda} \rangle_{\gamma_\lambda}\vollambda\leqslant 0
\end{eqnarray*}
where $s_{\lambda}$ is the connection one-form 
of $S_{\lambda}$, defined as  $\bm{s}_{\lambda} (X) \defi - \frac{1}{2}\langle k,\nabla_{X}l\rangle_{\gamma_\lambda}$
for any vector field $X$ tangent to $S_{\lambda}$.
\end{Lem}
\begin{proof}
Since the variation vector of the flow $\{ S_{\lambda} \}$ is $-k$ we need
to calculate
\begin{eqnarray}
\label{DerivativeMb}
\frac{d M_b(\lambda)}{d \lambda} & = &  
- 8 \pi + 
\int_{S_\lambda} \delta_{-k} \left ( \theta_l(\lambda) \right )  \volunitdos+
\int_{S_\lambda}\theta_l(\lambda)(\delta_{-k}\volunitdos) \\
 & = &
- 8 \pi + 
\int_{S_\lambda}\Big{(} \delta_{-k} (\theta_l(\lambda)) - \theta_k(\lambda)
\theta_l (\lambda) \Big{)} \volunitdos,
\end{eqnarray}
where $\delta_{-k}$ stands for geometric variation along $-k$ and 
we have used  the first 
variation of volume $\delta{_k}(\vollambda)=\theta_k(\lambda) \vollambda$
(see e.g \cite{Jost}) in the second equality.
The first variation of the null
expansion $\theta_l$ is  standard and can be found in many places
(see e.g. (2.23) in \cite{BoothFairhurst2007} 
where we need to set $\kappa=0$ because our null vector
$-k$ is geodesic and affinely parametrized, or Lemma 3.1 in \cite{AMS2} with $a=0$ for the same
reason)
\begin{equation}
\delta_{-k} \left ( \theta_l(\lambda) \right ) = \mbox{Scal} (S_{\lambda})+ \theta_l (\lambda)
\theta_k (\lambda) - 2\langle s_{\lambda},s_{\lambda} \rangle_{\gamma_\lambda}
+ 2\mbox{div}_{\gamma_\lambda} s_{\lambda}, \label{variation}
\end{equation}
where $\mbox{Scal} (S_{\lambda})$ is the scalar curvature of
$(S_\lambda,\gamma_{\lambda})$. The surfaces $S_{\lambda}$ are topologically
spheres, so the Gauss-Bonnet theorem gives
\begin{equation*}
\int_{S_\lambda} \mbox{Scal}(S_{\lambda})\vollambda=8\pi.
\end{equation*}
Inserting (\ref{variation}) 
in (\ref{DerivativeMb})  and using the Gauss-Bonnet theorem proves the 
Lemma.
\end{proof}

\begin{Tma}[{\bf Class of surfaces where the Penrose
inequality in $\mathcal{M}^{1,3}$  holds}]
\label{tsolution}
Let $(S,\gamma)$ be a spacetime strictly convex surface in $(\mathcal{M}^{1,3},\eta)$.
With the same assumptions and notation as in Theorem 
$\ref{T2}$, let $h$ be the support function of $\widehat{S}_0$ 
as a hypersurface of Euclidean space and 
$\tau =t |_{S} -t_0$. If these two functions satisfy the inequality
\begin{equation}
\label{C3}
4\pi\int_{\mathbb{S}^2}\big{(}(\triangle_{\metesf} h)^2+2h\triangle_{\metesf}
h\big{)}\volunitdos\geqslant
4\pi\int_{\mathbb{S}^2}u^2\volunitdos- \left (\int_{\mathbb{S}^2}u\volunitdos \right  )^2  
\end{equation}
where $u \defi\triangle_{\metesf} h+2(h-\tau)$, then the Penrose inequality (\ref{B2})
holds for $S$.
\end{Tma}

\begin{proof}
In analogy with the definition of $M_{b} (\lambda)$ (\ref{Bergqvist}), we define a
function $D(\lambda)$  by
\begin{eqnarray}
D(\lambda) \defi \sqrt{16 \pi |S_{\lambda}|} - 8 \pi \lambda.
\label{Dlambda}
\end{eqnarray}
With this definition, the Penrose inequality 
(\ref{B2}) for $S_{\lambda}$ can be written in the form
\begin{eqnarray*}
M_b (\lambda ) \geq D(\lambda).
\end{eqnarray*}
Our aim is to prove $M_b (\lambda=0) \geq D(\lambda=0)$.
Since Bergqvist's Lemma \ref{BergLem} ensures that $M_b(\lambda)$ is monotonically decreasing
in $\lambda$, the idea of the proof is to study the monotonicity properties
of $D(\lambda)$ together with the limiting behaviour of both functions 
when $\lambda \rightarrow \infty$  in order to see if sufficient conditions
can be obtained so that $M_b (\lambda=0) \geq D(\lambda=0)$ holds.

Let us start with the limit of $M_b(\lambda)$ at infinity. We want to exploit the fact that
we obtained in (\ref{Totalthetal})  a general
expression for the total integral of the outer null expansion $\theta_l$ on 
any spacetime convex surface $S$, in particular for $S_{\lambda}$. We need to
determine the support and time height function of $S_{\lambda}$. Although it is not the only
natural possibility, a convenient choice is to 
fix one  hyperplane $\Sigma_{t_0}$  and project all surfaces
$S_{\lambda}$ along $\Omega$ onto $\Sigma_{t_0}$. This procedure has the
advantage that $\widehat{S}_0$ is the same surface for all $S_{\lambda}$ and
hence that the support function $h$ is independent of $\lambda$.
With this choice, the time height function $\tau_{\lambda}$ of $S_{\lambda}$ 
is 
\begin{eqnarray*}
\tau_{\lambda} = t |_{S_{\lambda}} - t_0 = \sigma |_{S_{\lambda}} =  \tau - \lambda.
\end{eqnarray*}
Inserting these functions in (\ref{Totalthetal}) we find
\begin{align*}
& M_b(\lambda) =  \\
& = \int_{\mathbb{S}^2}\left (
\triangle_{\metesf} h+2(h-\tau) + 
\left ( \triangle_{\metesf} h+2(h-\tau + \lambda)  \right )
\left ( W_1(\lambda) | \conesf \tau|^2_{\metesf} - W_2(\lambda) (\conesf^A \conesf^B h) \conesf_A \tau \conesf_B \tau  \right )
\right )  \volunitdos,
\end{align*}
where $W_1 (\lambda)$ and $W_2(\lambda)$ are  obtained from (\ref{W1})-(\ref{W2}) after substituting $\tau 
\rightarrow \tau - \lambda$. Since $W_1(\lambda)$ and $W_2(\lambda)$ vanish as $\lambda^{-2}$ when $\lambda \rightarrow 
\infty$ the limit of $M_b(\lambda)$ is simply
\begin{eqnarray}
\lim_{\lambda\to\infty}M_b(\lambda)= \int_{\mathbb{S}^2}(\triangle_{\metesf} h+2(h-\tau))\volunitdos. \label{limMb}
\end{eqnarray}
Regarding $D(\lambda)$, we substitute $\tau \rightarrow \tau_{\lambda}$ in (\ref{A11}) to obtain
\begin{eqnarray}
|S_{\lambda}| = 
\int_{\mathbb{S}^2}\left ((h-\tau+\lambda)^2+
\triangle_{\metesf} h \left (\frac{h}{2}-\tau+\lambda \right ) \right ) \volunitdos, 
\label{area}
\end{eqnarray}
so that
\begin{eqnarray*}
D(\lambda) = \sqrt{16 \pi 
\int_{\mathbb{S}^2}\left ((h-\tau+\lambda)^2+
\triangle_{\metesf} h \left (\frac{h}{2}-\tau+\lambda \right ) \right ) \volunitdos}
- 8 \pi \lambda.
\end{eqnarray*}
It is straightforward to check that the limit of this expression at
infinity is
\begin{eqnarray*}
\lim_{\lambda \rightarrow \infty} D(\lambda) = \int_{\mathbb{S}^2}\left ( \triangle_{\metesf} h+2(h-\tau)\right )\volunitdos,
\end{eqnarray*} 
which coincides with the limit of $M_b(\lambda)$ obtained in (\ref{limMb}). Since $M_b(\lambda)$ is
monotonically decreasing, and $M_b(\lambda)$ coincides with $D(\lambda)$ at infinity, a sufficient
condition for the validity of $M_b(\lambda = 0) \geq D (\lambda=0)$ is that $D(\lambda)$ is monotonically
increasing. From the definition (\ref{Dlambda}) it follows
\begin{eqnarray}
\label{derivarea}
\frac{d D(\lambda)}{d\lambda} =  \sqrt{\frac{4 \pi}{|S_{\lambda}|}} \left ( \frac{d |S_{\lambda}|}{d\lambda}
- \sqrt{16  \pi |S_{\lambda}|} \right).
\end{eqnarray}
It only remains to find out under which conditions the right-hand side of $\eqref{derivarea}$ 
is non-negative. Since $\frac{d |S_{\lambda}|}{d \lambda} \geq 0$  (because $\theta_k \leq 0$ on $\Omega$
and $\frac{d \bm{\eta_{S_{\lambda}}}}{d \lambda} = - \theta_k (\lambda) 
\bm{\eta_{S_{\lambda}}}$), this is equivalent to
\begin{eqnarray}
\left ( \frac{d |S_{\lambda}|}{d\lambda} \right )^2 - 16 \pi |S_{\lambda}| \geq 0. \label{condition}
\end{eqnarray}
 It is now a matter of simple algebra to show that
(\ref{condition}) is equivalent to (\ref{C3}) with the definition  $u \defi \triangle_{\metesf} h+2(h-\tau)$.
\end{proof}

Theorem \ref{tsolution} gives a class of spacetime strictly convex surfaces in the Minkowski spacetime
for which the Penrose inequality holds.  An important question regarding this result is how
large is the class of surfaces covered by the theorem. 
Since inequality (\ref{C3}) is quadratic in $h, u$ and its derivatives,
a natural strategy is to expand these functions in terms  of spherical harmonics and
to rewrite (\ref{C3}) as an inequality for the coefficients of these
expansions. 

Let $r \in \mathbb{N} \cup {0}$ and 
$Y^r_m$ ($m=-r,\cdots r)$ be $2r +1$ linearly independent eigenfunctions of the spherical Laplacian
with eigenvalue $-r(r+1)$, i.e. $\triangle_{\metesf} Y^r_m = - r (r+1) Y^r_{m}$. Without loss
of generality we assume that they form an orthonormal basis of $L^2(\mathbb{S}^2)$, 
i.e. $\int_{\mathbb{S}^2} Y^r_{i} Y^s_{j} \volunitdos = \delta^{rs}\delta_{ij}$. Any smooth function $f$ 
on the sphere can be decomposed in this basis as
\begin{equation*}
f =\sum_{r=0}^\infty a_r\cdot Y^r,
\end{equation*}
where here and in the following we use the notation  $a_r\cdot Y^r \defi \sum_{m=-r}^r a_r^m Y^r_m$. Similarly
we write $a_r^2  \defi \sum_{m=-r}^r (a_r^m)^2$. The following
Corollary identifies the class of surfaces covered in Theorem \ref{tsolution} in terms of the spherical harmonic decompositions of $h$ and $u$.

\begin{Cor}
With the notation of Theorem \ref{tsolution}, let us expand the functions $h$ and $u$ in terms of spherical harmonics
as 
\begin{equation}
h=\sum_{r=0}^\infty a_r\cdot Y^r,  \quad u=\sum_{r=0}^\infty b_r\cdot Y^r.
\label{decom}
\end{equation}
If the coefficients satisfy the inequality
\begin{equation}
\label{esfcondition}
\sum_{r=2}^\infty a_r^2 r(r+1)(r-1)(r+2)\geqslant \sum_{r=1}^\infty b_r^2,
\end{equation}
then the Penrose inequality holds for the spacetime convex surface $S$ defined by the support function $h$ and
the time height function $\tau = h - \frac{1}{2} \triangle_{\metesf} h - \frac{u}{2}$.
\end{Cor}
\begin{proof}
The orthogonality relations of the spherical harmonics imply
\begin{equation*}
\int_{\mathbb{S}^2}(\triangle_{\metesf} h)^2\volunitdos=\sum_{r=0}^\infty a_r^2 r^2(r+1)^2, \quad \quad
\int_{\mathbb{S}^2}h\triangle_{\metesf} h\volunitdos= -\sum_{r=0}^\infty r(r+1)a_r^2,
\end{equation*}
so that  the left-hand side of  $\eqref{C3}$ reads 
\begin{equation}
\label{A17}
4\pi\int_{\mathbb{S}^2} \left ( (\triangle_{\metesf} h)^2+2h\triangle_{\metesf}
h \right ) \volunitdos=4\pi\sum_{r=2}^\infty a_r^2
r(r-1)(r+1)(r+2).
\end{equation}
On the other hand, the spherical harmonic decomposition of $u$ implies
$\int_{\mathbb{S}^2}u\volunitdos= \sqrt{4\pi}b^0_0$ and
 \begin{equation}
\label{A17BIS}
4\pi\int_{\mathbb{S}^2}u^2\volunitdos- \left (\int_{\mathbb{S}^2}u\volunitdos \right )^2=4\pi\sum_{r=0}^\infty
b_r^2-(\sqrt{4\pi}b^0_0)^2=4\pi\sum_{r=1}^\infty b_r^2.
\end{equation}
Using $\eqref{A17}$ and $\eqref{A17BIS}$, we obtain $\eqref{esfcondition}$, as claimed (we note in 
passing that (\ref{A17}) and (\ref{A17BIS}) imply that both sides in inequality 
(\ref{C3}) are non-negative).

\end{proof}

{\bf Remark 1.} In Theorem \ref{tsolution} we have shown that the Penrose inequality in the spherical case holds as 
a consequence of the Beckner inequality (or as a consequence of the Sobolev inequality in $\mathbb{R}^m$
in the case of four spacetime dimensions \cite{Tod1985}). It is interesting to see how does the spherical case
fit into the class of functions covered in Theorem \ref{tsolution}. It is well-known (and easy to proof) that the
support function of a sphere is either a constant (if the origin of Euclidean space coincides with the
center of the sphere) or a linear combination of $r=0,1$ spherical harmonics (when the sphere is displaced from the origin).
In either case, the left-hand side of (\ref{esfcondition}) is identically vanishing, so that the inequality can only hold 
if the right-hand side also vanishes. This forces $u=\mbox{const.}$ and hence $\tau = \mbox{const.}$ too.
We see that that the only ``spherical case'' included in Theorem \ref{tsolution} is
when the surface $S$ itself is spherically symmetric, which is a trivial case.
Thus, in some sense, the cases covered by Beckner's inequality (which
is essentially analytic in nature) and the cases covered by the
geometric flow used in Theorem \ref{tsolution} are mutually exclusive. This seems to indicate that any attempt
of proving the Penrose inequality for spacetime convex surfaces in the general case most likely needs some sort of combination
of  both ingredients and almost surely a combination of analytic and geometric arguments.

\vspace{5mm}

{\bf Remark 2.} The other case where the Penrose inequality in Minkowski was known to hold involves surfaces lying in
a constant time hyperplane. It is also natural to see how does this case fit into the class of surfaces covered
by Theorem \ref{tsolution}. In this situation we have $\tau=0$ and hence $u= \triangle_{\metesf} h+2h$. Inserting this function in 
(\ref{C3}), this inequality becomes
\begin{eqnarray}
\left (\int_{\mathbb{S}^2}h\volunitdos\right )^2\geqslant
  4\pi\int_{\mathbb{S}^2}\big{(}h^2+\frac{1}{2}h\triangle_{\metesf}
  h\big{)}\volunitdos,   \label{Mink2}
\end{eqnarray} 
which is exactly the Minkowski inequality for 2-dimensional
euclidean surfaces in terms of the support function (see formula (\ref{minkineq})). Since
the Minkowski inequality is true, it follows that the class of surfaces covered
by Theorem \ref{tsolution} includes the case of convex surfaces lying on constant time hyperplanes (incidentally,
it is immediate to prove directly the validity of (\ref{Mink2}) by using the spherical harmonic decomposition
$h=\sum_{r=0}^\infty a_r\cdot Y^r$). 

\vspace{5mm}


We finish this section, and the paper, with a particular case of Theorem \ref{tsolution} where the inequality
(\ref{C3}) can be interpreted nicely in terms of the geometry of the projected surface $\widehat{S}_0$
and of the height function $\tau$ of $S$.

\begin{Cor}
Let $\widehat{S}_0$ be a strictly convex surface embedded in a hyperplane $\Sigma_{t_0}$ and let $\Omega$
the spacetime convex null hypersurface containing $\widehat{S}_0$. The Penrose inequality holds for
any surface $S$ embedded in $\Omega$ and defined by a function $\tau = t |_{S} - t_0$ of the form
\begin{eqnarray*}
\tau = 2 \alpha\frac{\hat{J} (\widehat{S}_0)}{\mbox{Scal}(\widehat{S}_0)}-\beta
\end{eqnarray*}
where $\alpha \in [0,1]$, $\beta \in \mathbb{R}$ and $\hat{J}(\widehat{S}_0)$, $\mbox{Scal} (\widehat{S}_0)$  are, respectively, the mean
curvature and scalar curvature of $\widehat{S}_0$ as a hypersurface of Euclidean space.
\end{Cor}

\begin{proof}
For surfaces in $\mathbb{R}^3$, the scalar curvature can be written as
 $\mbox{Scal} (\widehat{S}_0) = 2 \kappa_1 \kappa_2$
where $\kappa_1$ and $\kappa_2$ are the principal curvatures of $\widehat{S}_0$ (hence positive everywhere since $\widehat{S}_0$ is strictly convex).
Since $\hat{J} = \kappa_1 + \kappa_2$
it follows 
\begin{eqnarray*}
\tau = 2 \alpha\frac{\hat{J} (\widehat{S}_0)}{\mbox{Scal}(\widehat{S}_0)}-\beta = 
\alpha \left ( \frac{1}{\kappa_1} + \frac{1}{\kappa_2} \right ) - \beta = \alpha \,
\mbox{tr}({\bf K_0}^{-1}) - \beta = \alpha \, \mbox{tr} ({\bf B}) - \beta =
\alpha \left ( \triangle_{\metesf} h+2 h \right ) - \beta.
\end{eqnarray*}
As a consequence, the function $u=\triangle_{\metesf} h+2(h-\tau)$ takes the form
\begin{eqnarray*}
u = (1-2\alpha) \left ( \triangle_{\metesf} h+2 h \right ) + 2 \beta,
\end{eqnarray*}
which in terms of the coefficients in the expansion  $\eqref{decom}$ implies
\begin{eqnarray*}
b_r= - (1-2\alpha) (r+2) (r-1)  a_r \quad r\geqslant 1 
\end{eqnarray*}
Inserting this into (\ref{esfcondition}) we find that this inequality becomes
\begin{eqnarray*}
\sum_{r=2}^{\infty} a_r^2 (r+2)(r-1) 
\left [ \frac{}{} r (r+1) - (1- 2 \alpha)^2 (r+2)(r-1) \right ] \geqslant 0 
\end{eqnarray*}
Since $h$ is basically arbitrary (it is only restricted by the condition that it defines
a strictly convex surface) we need to impose that each term of the sum is non-negative.
This is achieved only if 
\begin{eqnarray}
(1-2\alpha)^2 \leqslant \frac{r(r+1)}{(r+2)(r-1)} \defi Z(r), \quad \quad \forall r \geq 2. \label{condition2}
\end{eqnarray}
Since the sequence $Z(r)$ is decreasing and its limit is $1$ we see that this inequality holds if 
$(1 - 2 \alpha )^2 \leqslant 1$, which is equivalent to $\alpha \in [0,1]$. Since $\alpha$ is restricted
to this range by hypothesis, (\ref{condition2}) holds true and the Corollary follows.
\end{proof}

\section{Acknowledgments}

M.M. is grateful to Lars Andersson, G\"oran Bergqvist, Jan Metzger,
Miguel S\'anchez Caja  and Jos\'e M.M. Senovilla for useful
discussions. Financial support under the projects FIS2009-07238 (Spanish MEC)
and P09-FQM-4496 (Junta de Andaluc\'{\i}a and FEDER funds) are acknowledged. A.S. acknowledges the Ph.D. grant AP2009-0063 (MEC).

\end{document}